\documentclass[aip, amsmath,amssymb, reprint]{revtex4-1}
\usepackage{graphicx}
\usepackage{bm}
\usepackage[hidelinks]{hyperref}
\usepackage[capitalise]{cleveref}
\usepackage[utf8]{inputenc}
\usepackage[T1]{fontenc}
\usepackage{booktabs}
\usepackage{xcolor}
\usepackage[caption=false]{subfig}
\DeclareMathOperator{\atantwo}{atan2}

\begin{document}

% \preprint{AIP/123-QED}

%%%%%%%%%%%%%%%%%%%%%%%%%%%%%%%%%%%%
%% Template of Physics of Plasmas %%
%%%%%%%%%%%%%%%%%%%%%%%%%%%%%%%%%%%%

\title{Simplified and Flexible Coils for Stellarators using Single-Stage Optimization}
\author{R. Jorge}
 \email{Second.Author@institution.edu.}
\affiliation{Department of Physics, University of Wisconsin-Madison, Madison, Wisconsin 53706, USA}
\author{A. Giuliani}
\affiliation{Flatiron Institute, 162 Fifth Avenue, New York, NY 10010, USA}
\author{J. Loizu}
\affiliation{École Polytechnique Fédérale de Lausanne (EPFL), Swiss Plasma Center (SPC), CH-1015 Lausanne,
Switzerland}
\begin{abstract}
Single-stage optimization, also known as combined plasma-coil algorithms or direct coil optimization, has recently emerged as a possible method to expedite the design of stellarator devices by including, in a single step, confinement, stability, and engineering constraints. In this work, we show how such frameworks allow us to find new designs in a streamlined manner, yielding a broad range of new configurations. Examples are shown for stellarators with a small number of coils and quasisymmetric stellarators with only one to three coils per half field period, with external trim coils, helical coils, and a single set of coils generating both a quasi-axisymmetric and a quasi-helical equilibrium.
\end{abstract}

\maketitle

\section{\label{sec:intro} Introduction}

Stellarators are a type of fusion device that is able to rely solely on externally placed coils to confine a high-temperature plasma.
In contrast with tokamaks, such devices are non-axisymmetric and create a set of closed nested magnetic flux surfaces without the need for an externally induced plasma current \cite{Helander2014}.
The design and fabrication of current-carrying coils required for confining plasmas are prone to cause many hurdles, especially stellarator coils that require low tolerances and possess large shear stresses.
Given the large space of possible magnetic field equilibria and coil shapes in stellarators, great efforts have been devoted to exploring this space and finding suitable fusion power plant designs.
This work is devoted to a particular way to search this design space, namely using single-stage optimization, also known as combined plasma-coil algorithms or direct stellarator coil optimization.

As pointed out in Ref. \!\![\onlinecite{Zhu2017a}], the use of direct coil optimization was the prevailing method to design stellarator coils, including the use of analytical expressions, e.g., the general winding law \cite{Harmeyer1985}, and free boundary computer codes.
Then, plasma boundary optimization methods such as the one developed by Nührenberg and Zille \cite{Nuhrenberg1988} separated the problem of designing a high-performance plasma from developing a suitable coil set, the prevailing method used today.
Most stellarator experiments of today were designed in this manner, with the first step being the optimization of the magnetic field equilibrium to achieve good performance at some level, e.g., energetic particle confinement and MHD stability, and the second step being the shaping of coils that reproduce such magnetic field equilibrium.

Pioneering work in the second step was performed by Ref. \!\![\onlinecite{Merkel1987}] using winding surfaces, where the external magnetic field is produced by a current distribution on a surface surrounding the plasma.
Later developments designing coils with winding surfaces include the extended NESCOIL code \cite{Drevlak1998}, ONSET \cite{Drevlak1999}, NESVD \cite{Pomphrey2001}, COILOPT \cite{Strickler2002}, COILOPT++ \cite{Brown2016}, REGCOIL \cite{Landreman2017}.
Then, Ref. \!\![\onlinecite{Zhu2017a}] introduced a new method to design coils independently of winding surfaces using the FOCUS code.
This is important as winding surfaces have a large number of degrees of freedom and often require an extra optimization step on top of coil optimization.
Variants of FOCUS include FOCUSADD \cite{McGreivy2022}, which include automatic differentiation to help find finite-build coils and FAMUS \cite{Zhu2020}, which uses a topology optimization method to include permanent magnets in the design.
The code used in this work is SIMSOPT \cite{Landreman2021b}, which employs a FOCUS-like method to design coils, using a mixture of analytical derivatives and automatic differentiation to speed up the computation.
SIMSOPT has been successfully used to find coils that accurately reproduce quasisymmetric equilibria \cite{Wechsung2022,Wiedman2023}, with different levels of coil complexity and coil sensitivities.

However, such a two-step approach may be limiting.
As the equilibrium obtained in stage 1 does not contain direct information on how the resulting coils may be, a stage 2 optimization can lead to coils that are either extremely complex or coils that do not reproduce the target magnetic field to a sufficient level to guarantee similar performance.
Therefore, a combined plasma-coil design, or single-stage optimization, is preferred (see Ref. \!\![\onlinecite{Henneberg2021}] for a thorough review and categorization of different plasma-coil optimization algorithms).
A pioneering approach was introduced in Ref. \!\![\onlinecite{Giuliani2022a}] where, using the PyPlasmaOpt software, a single-stage optimization procedure was developed to find coils that yield analytical quasisymmetric equilibria based on the near-axis expansion \cite{Landreman2019b,Jorge2020,Landreman2020a}.
Using such an approach, we note that Ref. \!\![\onlinecite{Lee2022}] was able to find flexible stellarator equilibria that support multiple configurations by carefully placing planar coils and increasing the number of terms in the objective function.
The methods employed here are the single-stage optimization method using fixed boundary equilibria introduced in Ref. \!\![\onlinecite{Jorge2023}], the direct coil optimization method introduced in Ref. \!\![\onlinecite{Giuliani2023a}], and a new guided coil method.

In this work, we explore different types of coils and equilibria designs that were either previously employed in the context of two-stage stellarator optimization and can now be improved upon using a combined plasma-coil approach or require a combined approach to achieve their goal.
Different goals are pursued, namely drastically reducing the number of coils by using one to three coils per half of a field period, placing coils away from the plasma (similar to the concept of trim coils), using helical coils instead of modular coils, and having a flexible configuration with multiple equilibria for the same coils.
While our main optimization target is the existence of a set of nested closed flux surfaces with a given set of coils, for many of the examples shown here, we seek additional confinement metrics such as aspect ratio, rotational transform, and quasisymmetry.
We show here that novel algorithms that combine plasma and coil optimization allow for an accelerated design of new configurations while targeting several physics and engineering constraints simultaneously.
We note that such algorithms only consider magnetic field equilibria in a vacuum and, therefore, these are the types of equilibria considered here.
The methods used to optimize each device are stated in \cref{sec:methods}.

This paper is organized as follows.
In \cref{sec:methods}, the numerical algorithm and set of equations used are described.
Then, the following set of cases is considered:
simplified configurations with a low number of coils in \cref{sec:simple},
quasisymmetric stellarators with a low number of coils in \cref{sec:qslowcoils},
quasi-axisymmetric stellarators with externally placed trim coils in \cref{sec:QAtrimcoil},
quasisymmetric with helical coils in \cref{sec:helicalsection},
flexible quasi-axisymmetric and quasi-helically symmetric configurations with a single set of coils \cref{sec:QAQHflexible}.
The conclusions follow.

\section{\label{sec:methods} Optimization Method}

The optimization is performed using the SIMSOPT code \cite{Landreman2021b}, which has the necessary modules to compute the objective functions used here for both stage 1 and stage 2 optimization, as well as its gradients.
In the following, we start by outlining the modeling of the coils and their respective Biot-Savart magnetic fields and then detail both the single-stage fixed boundary optimization method and the direct coil-plasma optimization algorithms employed here.
The scripts used to obtain the configurations in this manuscript, as well as coil data and figures, are present in \href{https://github.com/rogeriojorge/simple_coil_paper}{https://github.com/rogeriojorge/simple\_coil\_paper}.

\subsection{Modelling of Coils}
\label{sec:coils}

In this work, we consider coils to be single filaments of current modeled as a three-dimensional curve $\mathbf \Gamma(\theta)$ with $\theta$ a parametrization angle.
The examples here consider modular or helical coils that are placed in space without adhering to a winding surface (using a FOCUS-like approach \cite{Zhu2017a}), except the circular torus helical coil example shown in \cref{sec:helicalcircular}.
Regarding modular coils, each coil $i$ is modeled as a periodic function
%
% \begin{equation}
%     \mathbf \Gamma^{(i)}=[\Gamma_1^{(i)},\Gamma_2^{(i)},\Gamma_3^{(i)}]:[0,2\pi)\rightarrow \mathbb{R}^3,
% \end{equation}
%
$\mathbf \Gamma^{(i)}=[\Gamma_1^{(i)},\Gamma_2^{(i)},\Gamma_3^{(i)}]$
where
\begin{equation}
    \Gamma_j^{(i)}=c_{j,0}^{(i)}+\sum_{l=1}^{N_F}\left[c_{j,l}^{(i)}\cos(l \theta)+s_{j,l}^{(i)}\sin(l \theta)\right],
    \label{eq:coilexpression}
\end{equation}
yielding a total of $3\times(2N_F +1)$ degrees of freedom per coil.
The value of $N_F$ is stated within each example.
The degrees of freedom for the coil shapes are then
\begin{equation}
    \mathbf x_{\text{coils}}=[c_{j,l}^{(i)},s_{j,l}^{(i)}, I_i],
\end{equation}
with $I_i$ the current going through each coil.
While \cref{eq:coilexpression} is valid for modular non-planar coils, in the following, we also show examples of other types of coils, such as planar and helical coils.

We take advantage of the fact that stellarators possess different symmetries and only place a set of $N_C$ coils per half field-period $n_{\text{fp}}$, leading to a total of $2 \times n_{\text{fp}} \times N_c$ modular coils.
The remaining coils are set by the fact that every boundary in the examples shown here has stellarator symmetry and an integer number of toroidal field periods, labeled as $n_\text{nfp}$.
The parameterization used for the helical coils is detailed in \cref{sec:helicalsection}, with \cref{sec:helicalcircular} showcasing a helical coil on a circular torus winding surface and \cref{sec:helicalarbitrary} the more general case of a space curve.
The magnetic field $\mathbf B_{\text{ext}}$ of each coil is evaluated using the Biot-Savart law
\begin{equation}
    \mathbf B_{\text{ext}}(\overline{\mathbf x})=\frac{\mu_0}{4\pi}\sum_{i=1}^{2n_{\text{fp}}N_C} I_i \int_{\mathbf \Gamma_i}\frac{d \mathbf l_i\times \mathbf r}{r^3},
\end{equation}
where $d \mathbf l_i = \mathbf x'_i d\theta$ is the differential line element, $\theta$ is an angle-like coordinate that parameterizes the coil curve $\mathbf \Gamma_i$ and $\mathbf r=\overline{\mathbf x} - \mathbf x_i$ is the displacement vector between the evaluated point on the surface and the differential element.
Each coil is divided into an integer number of quadrature points, typically between 100 and 200, and the cost functions used to regularize the optimization problem and simplify the coils are the minimum distance between two coils, the length of each coil, their curvature and mean-squared curvature, as well as the linking number between coils.
More details on the form of each function can be found in Ref. \!\![\onlinecite{Wiedman2023}].

Finally, we define here variations of the parameterization introduced in \cref{eq:coilexpression} used to produce the results shown in this work.
We start with the parameterization of circular coils with radius $r_0$ whose position in space is given by its center $\mathbf r_{\text{center}}=[x_{\text{center}},y_{\text{center}},z_{\text{center}}]$ described in cartesian coordinates and the orientation unit vector $\mathbf{n}_c = [n_x,n_y,n_z]$ normal to the plane of the coil and centered at the coil center. The orientation vector $\mathbf{n}_c$ can be described with two spherical angles $(\theta_c,\varphi_c)$ via the relations
\begin{subequations}
\begin{align}
\theta_c &=\atantwo{(n_y,n_x)} \\ 
\varphi_c&=\atantwo{(\sqrt{n_x^2+n_y^2},n_z)}
\end{align}
\end{subequations}
where $\atantwo{(y,x)}$ is the two-argument arctangent function, also equal to the phase of the complex number $x+iy$. The degrees of freedom can therefore be written as
\begin{equation}
    \mathbf x_{\text{coils}}=[r_0, \mathbf r_{\text{center}},\theta_c, \varphi_c]
\end{equation}
which means there is a total of no more than 6 geometrical degrees of freedom per coil.

Additionally, we optimize planar non-circular coils.
These correspond to a curve that is restricted to lie in a plane, and the shape of the curve within the plane is represented by a Fourier series in polar coordinates, making the curve essentially two-dimensional.
In order to allow the stellarator to have a three-dimensional axis shape, the planar curve is not restricted to be vertical, but it is rotated in three dimensions using a quaternion, and finally, a translation is applied.
The Fourier series in polar coordinates is given
\begin{equation}
    r(\theta) = \sum_{l=0}^{{N_F}} r_{c,l}\cos(l\theta) + \sum_{l=1}^{{N_F}} r_{s,l}\sin(l \theta).
\end{equation}
and the rotation quaternion by \cite{Gallier2019}
\begin{align}
       \bf{q} &= [q_0,q_i,q_j,q_k]\nonumber\\
       &= [\cos(\theta / 2), \hat{x}\sin(\theta / 2), \hat{y}\sin(\theta / 2), \hat{z}\sin(\theta / 2)],
\end{align}
where $\theta$ is the counterclockwise rotation angle about a unit axis $(\hat{x},\hat{y},\hat{z})$.
The degrees of freedom can therefore be written as
\begin{equation}
    \mathbf x_{\text{coils}}=[r_{c,l}, r_{s,l}, q_0, q_i, q_j, q_k, \mathbf r_{\text{center}}].
\end{equation}
%
%where $\mathbf r_{\text{center}}=[x_{\text{center}},y_{\text{center}},z_{\text{center}}]$ is the center of the coil in cartesian coordinates.
%
As described in SIMSOPT's documentation \cite{Landreman2021b}, a quaternion is used for rotation rather than other methods for rotation to prevent gimbal locking during optimization.
The quaternion is normalized before being applied to prevent scaling of the curve.

We also optimize helical coils $\eta(\theta)$ using two different parameterizations.
The first one restricts the helical coil to lie on a circular torus, described by
\begin{equation}
    \eta(\theta) = n_{\text{fp}} \theta/l_0 + \sum_{k=0}^{N_F} A_k \cos(n_{\text{fp}} \theta k/l_0) + B_k \sin(n_{\text{fp}} \theta k/l_0),
\end{equation}
where $n_{\text{fp}}$ is the number of field periods and $l_0$ an integer.
The coefficients $A_k$ and $B_k$ are independent degrees of freedom and the sum goes from $k=0$ to a maximum order of $N_F$.
Such helical coils are restricted to wind around a circular torus with a major radius $R_0$ and a minor radius $r_0$ and have a position vector given by
\begin{align}
    x(\theta) &= R_0+r_0 \cos[\eta(\theta)] \cos(\theta),\\
    y(\theta) &= R_0+r_0 \cos[\eta(\theta)] \sin(\theta),\\
    z(\theta) &=-r_0 \sin[\eta(\theta)].
\end{align}
The second parameterization defines helical coils that are not restricted to lie on a particular surface but still possess $n_{\text{fp}}$-fold discrete rotational symmetry and stellarator symmetry.
These can be written in the following form
\begin{equation}
\begin{aligned}
    x(\theta) &= \cos(\theta) \eta_1(\theta) -\sin(\theta) \eta_2(\theta) ,\\
    y(\theta) &= \sin(\theta) \eta_1(\theta) +\cos(\theta) \eta_2(\theta) ,\\
    z(\theta) &= \sum_{k=1}^{N_F} z_{k}\sin( k n_{\text{fp}}\theta),
\end{aligned}
\end{equation}
where
\begin{equation}
\begin{aligned}
    \eta_1(\theta) &=  \eta_{10}+\sum_{k=1}^{N_F} \eta_{1k}\cos(k n_{\text{fp}}\theta),\\
    \eta_2(\theta) &= \sum_{k=1}^{N_F} \eta_{2k}\sin(k n_{\text{fp}}\theta),\\
\end{aligned}
\end{equation}
The degrees of freedom are given by $z_k, \eta_{10}, \eta_{1k}$ and $\eta_{2k}$.
This representation can also be used for non-stellarator symmetric helical coils by including additional symmetry-breaking harmonics.

\subsection{Single-Stage Optimization with Fixed Boundary Equilibria}
\label{sec:singlestage}

To calculate the magnetic field equilibrium, we use VMEC \cite{Hirshman1983}, which solves the static ideal magnetohydrodynamics (MHD) equation $\mathbf J \times \mathbf B = \nabla P$ where $\mu_0 \mathbf J = \nabla \times B$ is the plasma current density, $\mathbf B$ the equilibrium magnetic field satisfying $\nabla \cdot \mathbf B = 0$ and $P$ the plasma pressure.
VMEC assumes a toroidal equilibrium with nested magnetic flux surfaces and uses the steepest descent method to find a minimum in the potential energy $W$ resulting from an integral formulation of the ideal MHD equation.
VMEC is run in fixed boundary mode, where the outermost surface $S=[R(\vartheta,\phi)\cos(\phi), R(\vartheta,\phi)\sin(\phi), Z(\vartheta,\phi)]$, also called last closed flux surface, is fixed and used as a boundary condition. The boundary surface $S$ is specified by its Fourier amplitudes $\{\mathrm{RBC}_{m,n},\mathrm{ZBS}_{m,n}\}$ in cylindrical coordinates.
The degrees of freedom for the surface shapes are then
\begin{equation}
    \mathbf x_{\text{surface}}=\left[\text{RBC}_{m,n},\text{ZBS}_{m,n}\right],
\end{equation}
except $\text{RBC}_{0,0}$ which is kept at 1 to keep the overall dimensions of the device fixed.

In the single-stage optimization approach employed here, we minimize a combined objective (or cost) function, $J$, which consists of the sum of the objective functions associated with stage 1 and stage 2 optimization, $J_1$, and $J_2$, respectively.
The objective functions associated with the stage 1 optimization considered here are quasisymmetry, $f_{QS}$, defined as \cite{Landreman2022}
\begin{align}
    f_{\text{QS}} = &\sum_{s_j}\left<\left(\frac{1}{B^3}\left[(N-\iota M)\mathbf B \times \nabla B \cdot \nabla \psi\right.\right.\right.\nonumber\\
    &\left.\left.\left.-(MG+NI)\mathbf B \cdot \nabla B\right]\right)^2\right>,
\label{eq:fqs}
\end{align}
where $N$ is an integer describing the flavor of quasisymmetry, $G(\psi)$ is $\mu_0/(2\pi)$ times the poloidal current outside the surface, $I(\psi)$ is $\mu_0/(2\pi)$ times the toroidal current inside the surface, $\iota$ is the rotational transform and $\left<\dots\right>$ is a flux surface average.
The sum is over a set of flux surfaces $s_j=\psi_j/\psi_b$ where $\psi_b$ is the toroidal flux at the boundary and a uniform grid $0, 0.1, \dots, 1$ is used.
The quantities $\mathbf B \times \nabla B \cdot \nabla \psi$, $\mathbf B \cdot \nabla B$, $B$, $G$ and $I$ are computed using VMEC, while we set
\begin{equation}
    (M,N) = 
  \begin{cases}
    (1,0), & \text{for quasi-axisymmetry},\\
    (1,-1), & \text{for quasi-helical symmetry},
  \end{cases}
\label{eq:howqaqh}
\end{equation}
which are the two allowed flavors of quasisymmetry close to the magnetic axis \cite{Plunk2019}.

Additionally, we limit the aspect ratio $A$, defined as
\begin{equation}
    A = \frac{\texttt{Rmajor\_p}}{\texttt{Aminor\_p}}=\frac{V}{2\pi^2\texttt{Aminor\_p}^3}=\frac{V}{2\sqrt{\pi}\overline S^{3/2}},
\end{equation}
where $\overline S = (2\pi)^{-1}\int_0^{2\pi}d\phi S(\phi)$ is the toroidal average of the area $S(\phi)$ of the outer surface's cross section in the $R-Z$ plane and $V$ is the volume of the outer surface \cite{Landreman2019b}, by adding a target aspect ratio to $J_1$ of the form $f_A=(A-A_\text{target})^2$.
In some examples, a target rotational transform is added to $J_1$ in the form $f_\iota=(\iota-\iota_\text{target})^2$.
Finally, when quasisymmetry is not considered, we add a target mirror ratio $\Delta$ defined as
\begin{equation}
    \Delta = \frac{B_\text{max}-B_\text{min}}{B_\text{max}+B_\text{min}}.
\end{equation}
using the objective function $f_\Delta=(\Delta-\Delta_{\text{target}})^2$.

The objective function used to link the plasma and the coils is the quadratic flux quantity $f_{\text{QF}}$, given by
\begin{equation}
    f_{\text{QF}}=\int_S \left(\frac{\mathbf B_{\text{ext}} \cdot \mathbf n}{|\mathbf B_{\text{ext}}|}\right)^2 dS,
\label{eq:quadraticflux}
\end{equation}
where $\mathbf B_{\text{ext}}=\mathbf B_{\text{ext}}(\mathbf x_{\text{coils}})$, $\mathbf n = \mathbf n(S)$ and $S=S(\mathbf x_{\text{surface}})$.
If the field induced by the coils $\mathbf B_{\text{ext}}$ coincides with the target equilibrium field $\mathbf B$, then $f_{\text{QF}}=0$.
The objective function associated with stage 2 optimization, $J_2$, is then the sum of the quadratic flux, $f_{\text{QF}}$, and the regularization terms described at the end of \cref{sec:coils}.

The optimization problem and total objective function $J$ can then be formulated as
\begin{subequations} \label{eq:opt}
\begin{align}
    \min_{\mathbf x_{\text{coils}},\mathbf x_{\text{surface}}} &J(\mathbf x_{\text{coils}},\mathbf x_{\text{surface}})=J_1+\omega_{\text{coils}}J_2, \label{eq:J}\\
    \text{subject to } &~\psi=\psi_0,~R_{\text{major}}=R_0,
\end{align}
\end{subequations}
with $\omega_{\text{coils}}$ a scalar weight associated with $J_2$.
The degrees of freedom varied during the optimization are $\mathbf x=(\mathbf x_{\text{surface}}, \mathbf x_{\text{coils}})$.
The objective function, its gradients, as well as its CPU parallelization, are carried out using the SIMSOPT code \cite{Landreman2021b}.
The constraint $\psi=\psi_0$ is handled by the use of VMEC in fixed-boundary mode, while the constraint $R_{\text{major}}=R_0$ is handled by removing the term $\mathrm{RBC}_{0,0}$ from the parameter space and setting it equal to one.
We minimize $J$ by employing the Broyden–Fletcher–Goldfarb–Shanno (BFGS) quasi-Newton algorithm \cite{Nocedal2006}.
This method stores the Jacobian of the objective function computed at previous points and uses the BFGS formula to approximate the Hessian using a generalized secant method.
As noted in \cite{Wechsung2022}, this is important for convergence as it helps to overcome the ill-posedness of the coil optimization problem.
The Jacobian $dJ/d\mathbf x=dJ_1/d\mathbf x+dJ_2/d\mathbf x$ is computed using a mix of numerical and analytical derivatives.
In particular, the Jacobian $dJ_1/d\mathbf x_{\text{surface}}$ is computed using forward finite differences, $dJ_1/d\mathbf x_{\text{coils}}=0$, and both $dJ_2/d\mathbf x_{\text{coils}}$ and $dJ_2/d\mathbf x_{\text{surface}}$ are computed analytically.
The initial condition is a circular cross-section torus, except for cases where we optimize for quasi-helical symmetry, where torsion of the magnetic axis is added to the initial condition via $\text{RBC}_{1,1}$ and $\text{ZBS}_{1,1}$.
The parameter space is expanded in a series of steps, with the surface shape refined at each step.
In step j = 1, 2, $\cdots$, the amplitudes RBC$_{m,n}$ and ZBS$_{m,n}$ with $m\le j$ and $|n|\le j$ are varied.

The assessment of the resulting fixed boundary equilibria stemming from the optimization is done in the following way.
First, the properties of each equilibrium that are included in $J_1$ are verified, such as quasisymmetry, aspect ratio, rotational transform, and mirror ratio.
Second, a set of field field lines is traced to verify that the obtained coils reproduce the target fixed boundary equilibrium.
The field line tracing is performed using SIMSOPT with a tolerance for the ordinary differential equation (ODE) solver of $10^{-14}$ and the Biot-Savart field is interpolated in 3D to speed up the computation.
The tracing is performed both inside and outside the obtained plasma boundary in order to assess if there are flux surfaces away from the fixed boundary surface $S$.
Third, the fixed boundary equilibrium is converted to Boozer coordinates using the BOOZ\_XFORM code \cite{Sanchez2000a} and the contours of constant magnetic field strength are plotted on different surfaces to verify its degree of quasisymmetry further.
Finally, a set of 3500 alpha particles with an energy of $3.5$ MeV are traced using the code SIMPLE \cite{Albert2020} starting isotropically on a surface at a radius of $s=0.25$ and the fraction of loss particles after a total simulation time of $0.01$s is found.

We now detail the optimization for a flexible device involving a single set of currents with multiple equilibria.
In this case, we target two types of quasisymmetry: quasi-axisymmetry and quasi-helical symmetry.
This is performed by adding to the objective function two quasisymmetry terms from \cref{eq:fqs}: one with $N=0$ (QA) and one with $N=-1$ (QH), as dictated by \cref{eq:howqaqh}.
Each quasisymmetry term has its own VMEC class in SIMSOPT, with its own value of aspect ratio and rotational transform.
For the stage 2 objective function, two quadratic flux quantities $f_{\text{QF}}$ from \cref{eq:quadraticflux} are added together, with $S$ corresponding to each VMEC surface.
The degrees of freedom are the coefficients for each one of the surfaces, namely $\left[\text{RBC}_{m,n},\text{ZBS}_{m,n}\right]$, as well as the coil coefficients $x_{\text{coils}}=[c_{j,l}^{(i)},s_{j,l}^{(i)}, I^{QA}_i, I^{QH}_i]$ where $I^{QA}_i$ and $I^{QH}_i$ are the coil currents for the QA and QH configurations, respectively.

\subsection{Guided coil optimization}
\label{sec:guided}

{If only vacuum fields possessing magnetic surfaces and rotational transform are sought, a fast and reduced version of the method described in Sec.~\ref{sec:singlestage} might be attempted. The idea is to use a simplified boundary in the equilibrium code  (VMEC or SPEC \cite{Hudson2012,Hudson2020} can be used) to guide the coils towards a configuration with rotational transform. One starts by reducing the number of degrees of freedom for the boundary to only a few. This is done by restricting the number of Fourier modes to those that can produce on-axis rotational transform according to a near-axis expansion prediction \cite{Mercier1964}. For example, one can select the modes $[\text{RBC}_{0,0}, \text{RBC}_{1,0}, \text{ZBS}_{1,0}, \text{RBC}_{0,n}, \text{ZBS}_{0,n}]$ for a given $n$ and such that they describe a boundary that has circular cross-section and a constant torsion. The optimization is then carried out for the objective function given in  Eq.~(\ref{eq:opt}) and with $J_1=f_{\iota}+f_{QF}$. In this way, the coils try to reproduce best a magnetic surface that ensures rotational transform and,  given the restrictions on the guiding boundary, a large volume of magnetic surfaces might be expected. At the end of the optimization, one can assess the quality of the vacuum field from the coils and extract a Quadratic-Flux-Minimizing (QFM) \cite{Dewar1994} surface that can be used to run a high-resolution equilibrium calculation to evaluate physical quantities of interest.
These are surfaces that minimize the quantity
\begin{equation}
    f_{QFM}=\frac{\int_0^{2\pi} \int_0^{2\pi/n_{\text{fp}}} (\mathbf B \cdot \mathbf n)^2 d \theta d \varphi}{\int_0^{2\pi} \int_0^{2\pi/n_{\text{fp}}} B^2 d \theta d \varphi} + \frac{(\text{Vol}_S-\text{Vol}_\text{QFM})^2}{2},
\label{eq:QFM}
\end{equation}
without constraints on the angles that parametrize the surface, with $\textbf{n}$ the unit vector normal to the surface $S$ that is being optimized, $\text{Vol}_S$ its enclosed volume and $\text{Vol}_\text{QFM}$ the target volume.
An example of the application of this method is illustrated in Sec.~\ref{sec:circular} for a stellarator with only circular coils.} 

\subsection{Direct Coil Optimization using Boozer residuals}

This approach is based on Ref. \!\![\onlinecite{Giuliani2023a}] and does not rely on equilibrium solvers to perform direct stellarator coil design in a vacuum field.  
The optimization algorithm comprises three coil design phases, wrapped in globalization subroutines.
The first phase directly optimizes stellarator coils for near-axis quasisymmetry \cite{Giuliani2022a}.  This optimization algorithm is quite robust but sometimes finds devices that have nested surfaces on a small volume. 
The second phase takes the coils from the previous phase and attempts to increase the volume of nested flux surfaces and heal generalized chaos and islands \cite{Giuliani2023}. 
This is done by finding a magnetic field generated by coils that minimizes
\begin{equation}\label{eq:residual}
        f_{\text{Boozer}} = \int\int \| \mathbf r \|^2~d\varphi ~d\theta.
\end{equation}
on a magnetic surface $\Sigma(\varphi, \theta)$, where the residual is
\begin{equation}
    \mathbf r := G \frac{\mathbf B}{\|\mathbf B\|}-\|\mathbf B\|\left( \frac{\partial \Sigma}{\partial \varphi} + \iota \frac{\partial \Sigma}{\partial \theta}\right).
\end{equation}
The residual in \cref{eq:residual} is small when $\Sigma$ is a magnetic surface parametrized in Boozer coordinates in the magnetic field $\mathbf B$.  The residual is large in the presence of islands and generalized chaos, which motivates minimizing \cref{eq:residual} during the second coil design phase.
The final phase polishes the coils for an accurate approximation of quasisymmetry \cite{Giuliani2022b}.  
This is done by decomposing the field strength on the magnetic surfaces into a quasisymmetric and non-quasisymmetric component
$$
B(\varphi, \theta) = B_{\text{QS}}(\theta-N\varphi) + B_{\text{non-QS}}(\varphi, \theta).
$$
Then, the non-quasisymmetric component of the field strength is minimized
$$
f_{QS} =  \int_S B_{\text{non-QS}}(\varphi, \theta)^2 ~dS.
$$
The entire workflow is wrapped in a globalization algorithm called TuRBO \cite{TuRBO}.

\section{\label{sec:simple} Simplified Stellarators with a Low Number of Coils}

{
In this section, we consider configurations with one coil per half field-period, $n_\text{nfp}$, leading to a total of $2 n_\text{nfp}$ coils.
As the goal is to obtain a simplified stellarator with a small number of coils, the objective function does not include a quasisymmetry target.
We consider three cases of increasing complexity: circular, planar, and non-planar coils. }

\subsection{Circular Coils}
\label{sec:circular}

\begin{figure}
    \centering
    \includegraphics[clip,trim=0.2cm 0.1cm 0.1cm 0.1cm,width=.41\textwidth]{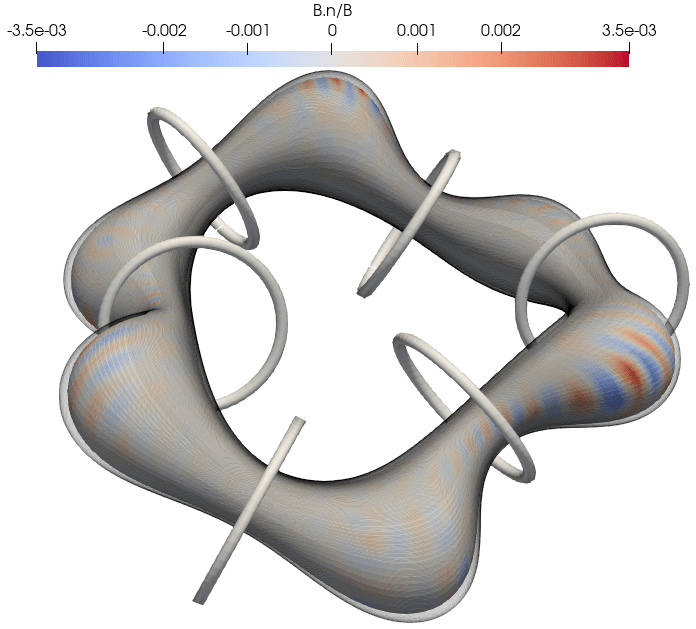}\\
    \includegraphics[clip,trim=0.0cm 0.0cm 0.0cm 0.0cm,width=.35\textwidth]{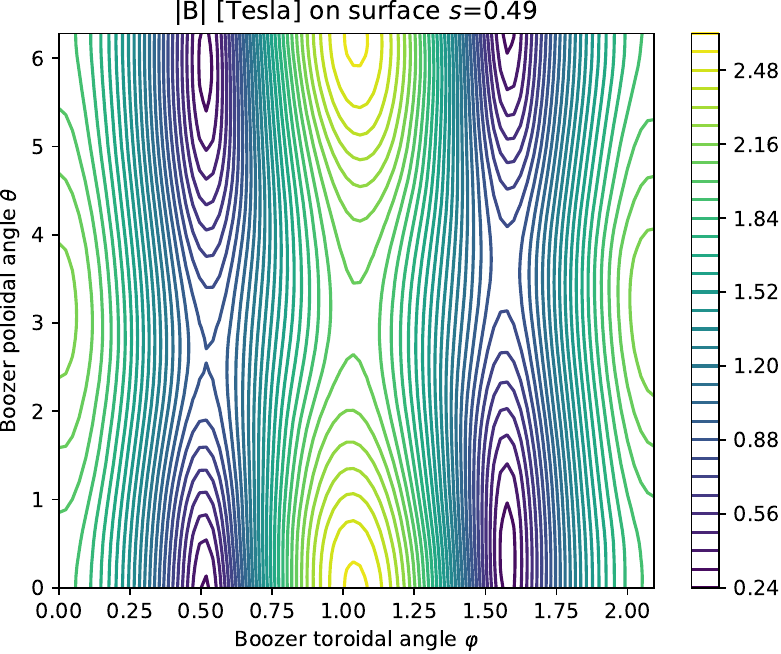}
    \includegraphics[clip,trim=2.3cm 6.8cm 11.6cm 0.3cm,width=.18\textwidth]{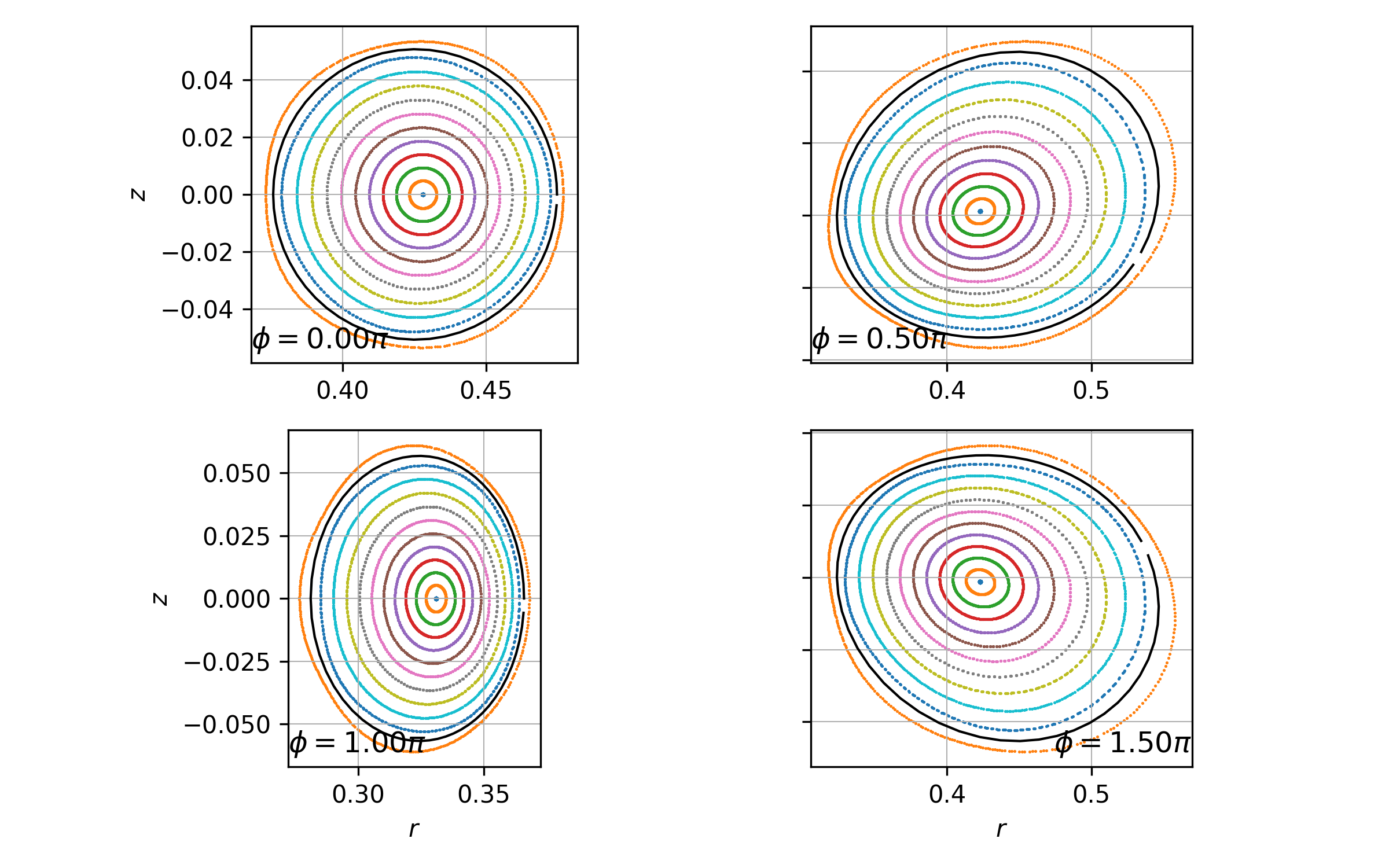}
    \includegraphics[clip,trim=0.1cm 11.9cm 22.7cm 0.7cm,width=.29\textwidth]{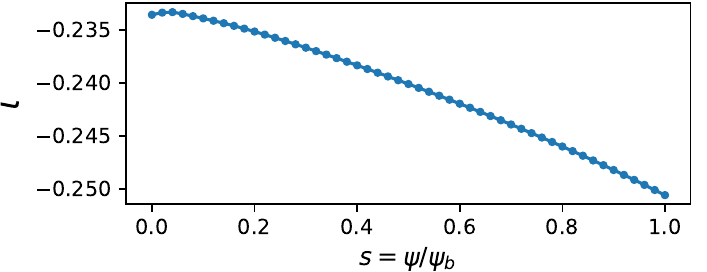}
    \caption{
    Three-field period stellarator with one circular coil per half field-period. Top: three-dimensional view of the fixed boundary magnetic field, the normalized error coil field $\mathbf B \cdot \mathbf n/B$, and magnetic field lines in black. Middle: Contours of constant magnetic field in Boozer coordinates at $s=0.5$. Bottom left: Field line tracing at $\phi=0$ and the plasma boundary in black. Bottom right: rotational transform.
    }
    \label{fig:nfp3_ncoils1_circular_params}
\end{figure}

{
We show in \cref{fig:nfp3_ncoils1_circular_params} a 3-field period stellarator with 1 circular coil per half field-period where the fixed boundary equilibrium has an aspect ratio of $A=5.2$.
The position of the 6 coils was obtained using the method described in \cref{sec:guided} with a target rotational transform $\iota_{\text{target}}=0.2$. During the optimization, the boundary of the equilibrium was constrained with the modes $[\text{RBC}_{0,0}=0.35, \text{RBC}_{1,0}=0.037, \text{ZBS}_{1,0}=-0.037]$ held fixed and the modes $[\text{RBC}_{0,3}, \text{ZBS}_{0,3}]$ free to vary. The radius of the coils, $r_0=0.15$, was fixed from the start. The current in all coils is the same. 
After the optimization, a QFM surface is extracted by optimizing a surface with a high Fourier resolution and using the objective function $f_{\text{QFM}}$ as given by \cref{eq:QFM}. A fixed boundary equilibrium is then obtained using the QFM surface as a boundary, which has a major radius of $\text{RBC}_{0,0}=0.4$.
Scaled to a major radius of one, the independent circular coil has a total length of 2.25, and five degrees of freedom (three for its center and two for its orientation).
The minimum separation between coils is 0.5 and the minimum distance between the coils and the plasma is 0.15.
The rotational transform profile is approximately constant of $\iota \simeq 0.24$ with a mirror ratio $\Delta=0.64$.
The normalized field error $\mathbf B \cdot \mathbf n/B$ with $\mathbf B$ the coil magnetic field and $\mathbf n$ the boundary normal vector is at most 0.35\%, leading to an extremely good agreement between the fixed boundary VMEC equilibria and the Poincaré sections, as evidenced in \cref{fig:nfp3_ncoils1_circular_params}.
}
We note that the configurations in the remaining sections were only constrained to a major radius of 1 meter ($\text{RBC}_{0,0}=1.0$)

\subsection{Planar Non-Circular Coils}

\begin{figure}
    \centering
    \includegraphics[clip,trim=3.2cm 2.1cm 3.9cm 1.5cm,width=.41\textwidth]{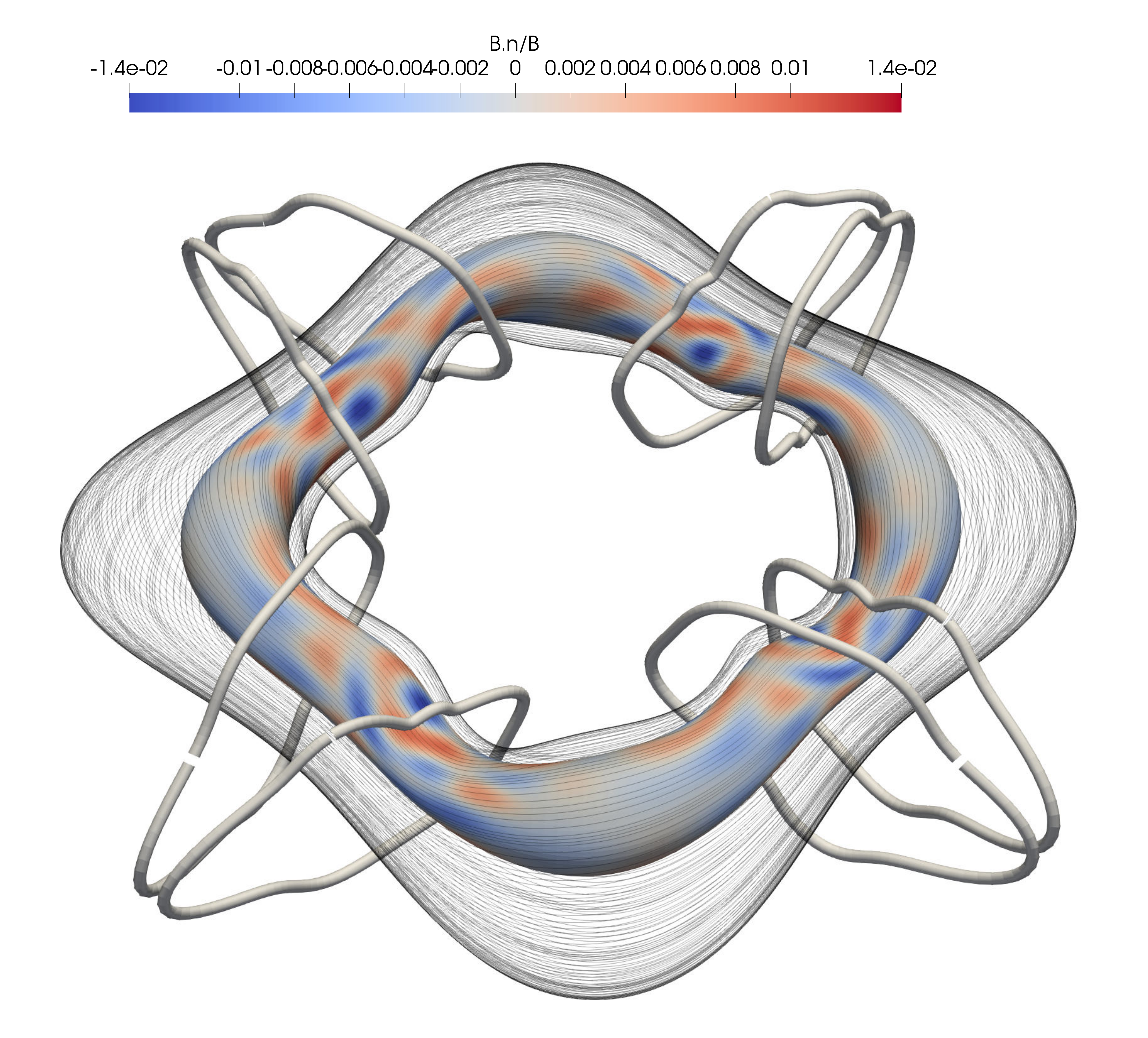}\\
    \includegraphics[clip,trim=0.0cm 0.0cm 0.0cm 0.0cm,width=.35\textwidth]{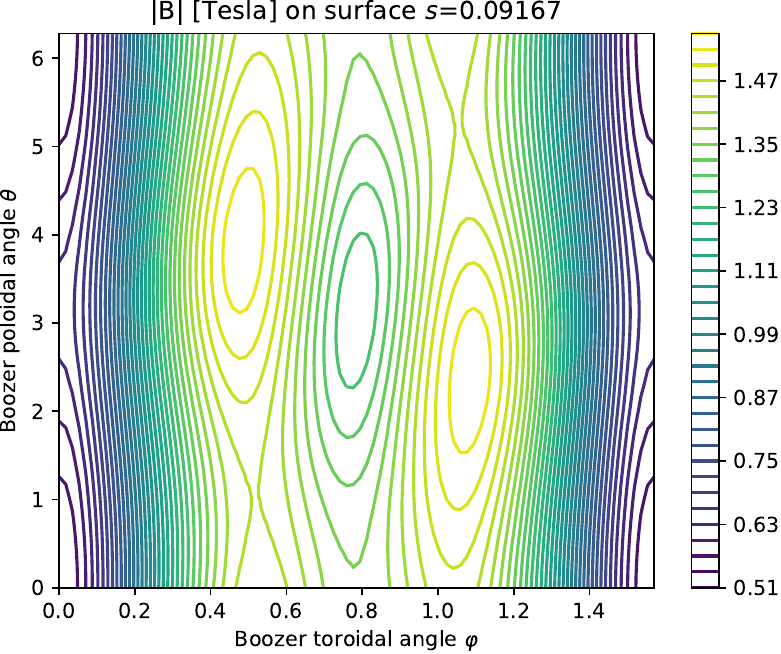}
    \includegraphics[clip,trim=1.0cm 6.3cm 9.5cm 0.4cm,width=.22\textwidth]{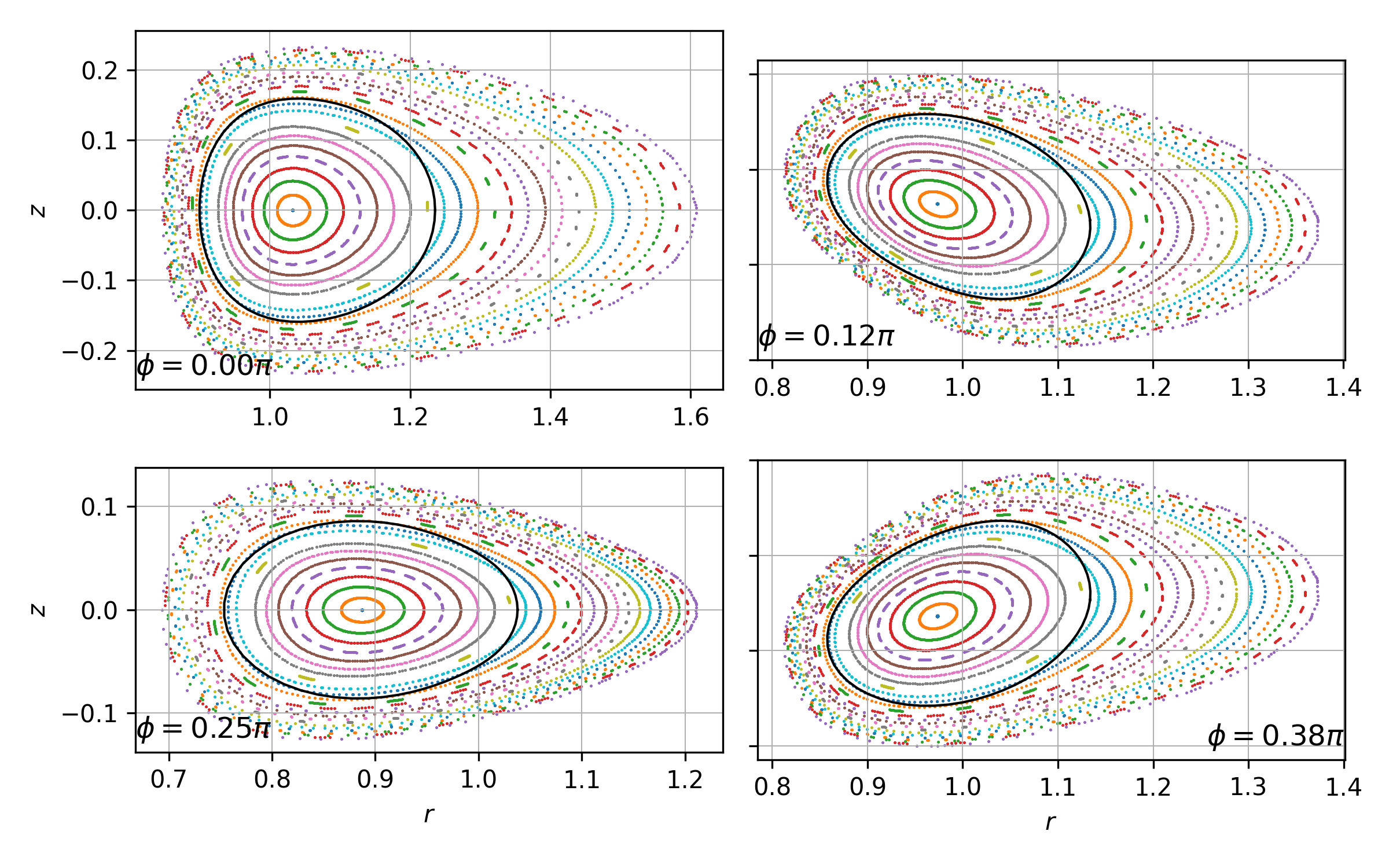}
    \includegraphics[clip,trim=0.1cm 11.9cm 22.9cm 0.7cm,width=.25\textwidth]{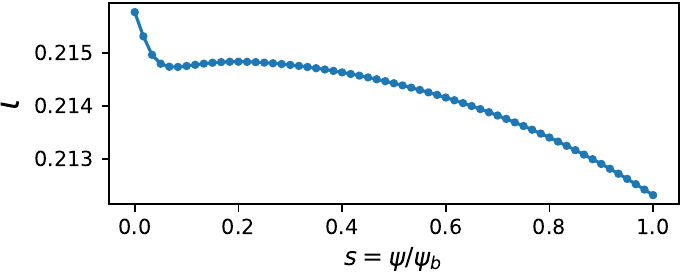}
    \caption{
    Four-field period stellarator with one planar coil per half field-period. Top: three-dimensional view of the fixed boundary magnetic field, the normalized error coil field $\mathbf B \cdot \mathbf n/B$, and magnetic field lines in black. Middle: Contours of constant magnetic field in Boozer coordinates at $s=0.1$. Bottom left: Field line tracing at $\phi=0$ and the plasma boundary (in black). Bottom right: rotational transform.
    }
    \label{fig:nfp4_ncoils1_planar_params}
\end{figure}

We show in \cref{fig:nfp4_ncoils1_planar_params} a 4-field period stellarator with 1 coil per half field-period where the fixed boundary equilibrium has an aspect ratio of $A=7.9$.
The independent coil has a total length of 3.3, a maximum curvature of 25.5, a mean-squared curvature of 22.0, and twenty degrees of freedom.
The minimum separation between coils is 0.1 and the minimum distance between the coils and the plasma is 0.13.
The rotational transform profile is approximately constant of $\iota \simeq 0.21$ with a mirror ratio $\Delta=0.45$.
The normalized field error on the boundary is at most 1.4\%, leading to some deviation between the fixed boundary VMEC equilibria and the Poincaré sections, as evidenced in \cref{fig:nfp4_ncoils1_planar_params}.
Nevertheless, we note that such a configuration has a large volume of flux surfaces that extends beyond the last closed surface computed using the single-stage approach.

\subsection{Non-Planar Coils}

\begin{figure}
    \centering
    \includegraphics[clip,trim=1.2cm 1.5cm 0.4cm 1.5cm,width=.41\textwidth]{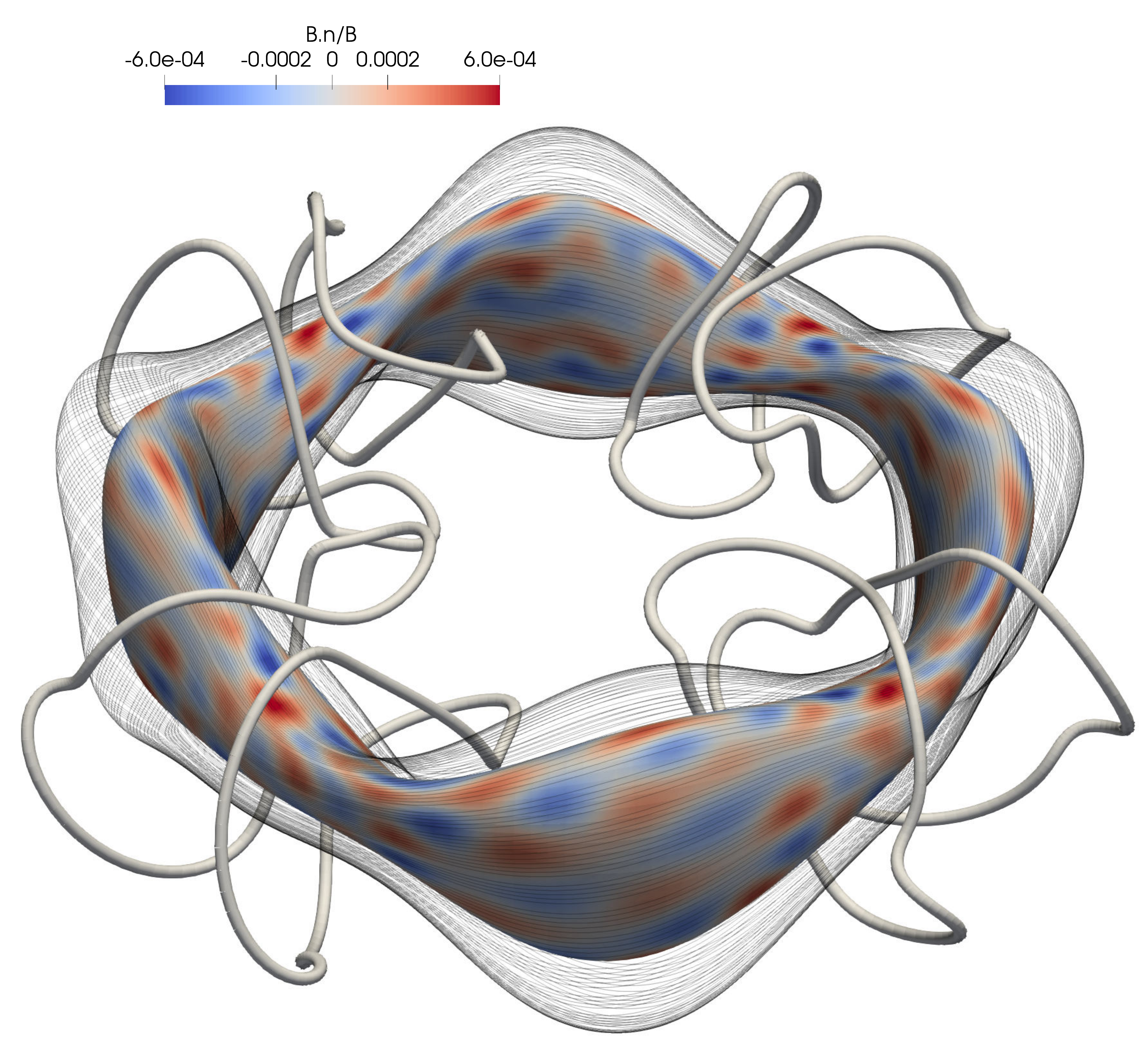}\\
    \includegraphics[clip,trim=4.5cm 6.5cm 12.7cm 0.3cm,width=.14\textwidth]{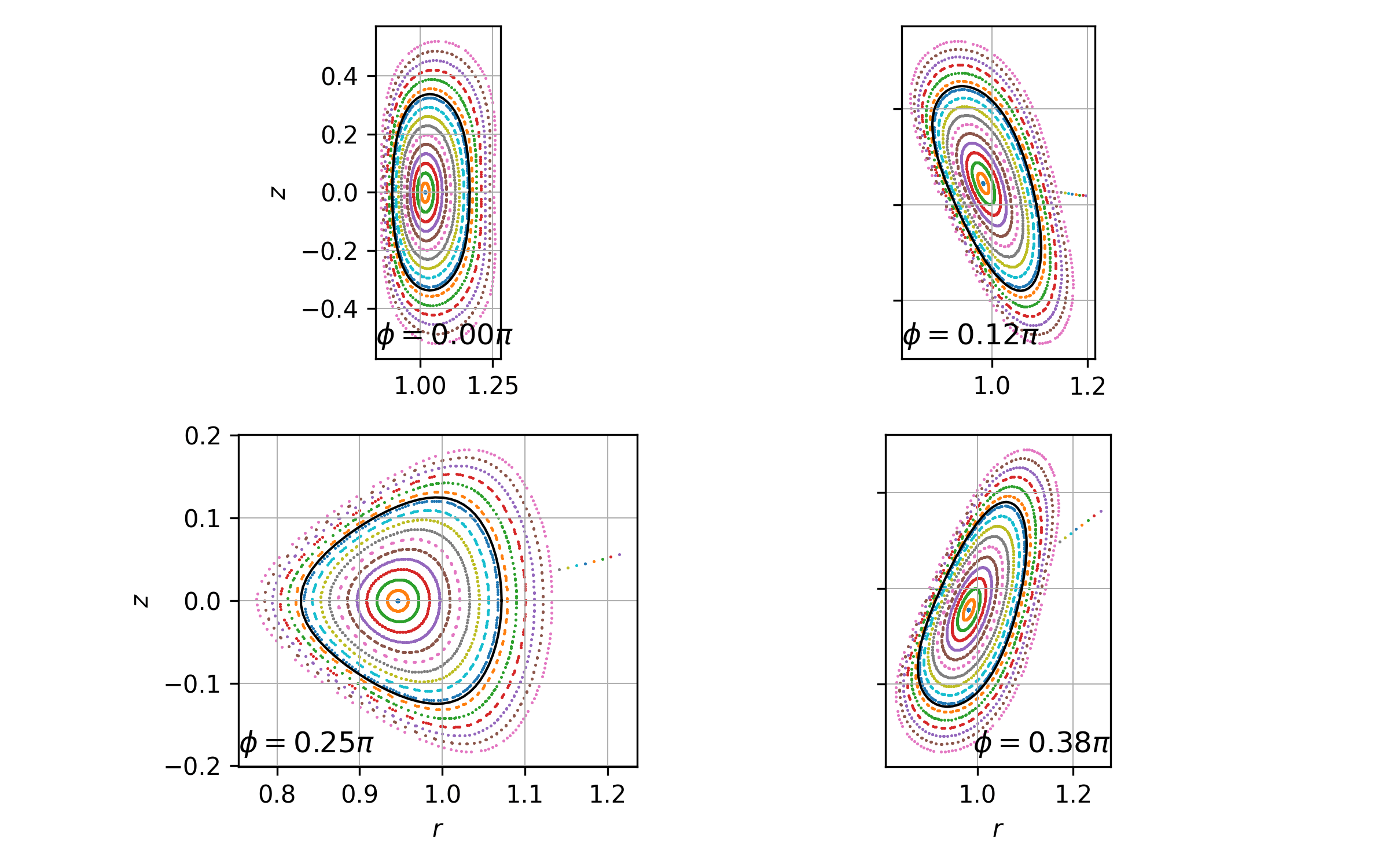}
    \includegraphics[clip,trim=0.0cm 0.0cm 0.0cm 0.0cm,width=.33\textwidth]{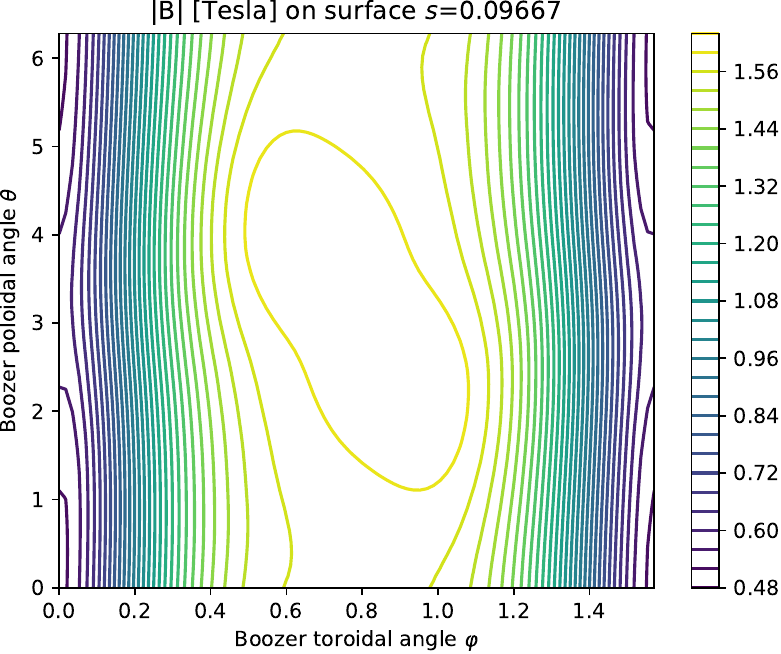}
    \includegraphics[clip,trim=0.1cm 11.9cm 23.0cm 0.7cm,width=.30\textwidth]{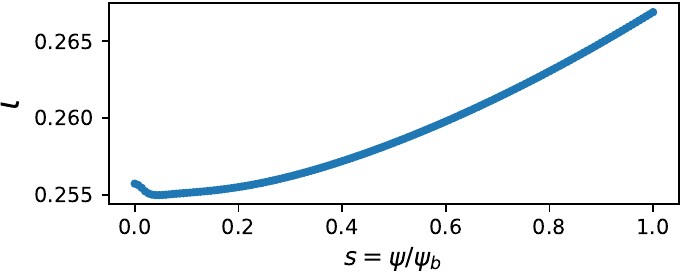}
    \caption{
    Four-field period stellarator with one non-planar coil per half field-period. Top: three-dimensional view of the fixed boundary magnetic field, the normalized error coil field $\mathbf B \cdot \mathbf n/B$, and magnetic field lines in black. Middle left: Field line tracing at $\phi=0$ and the plasma boundary (in black). Middle right: Contours of constant magnetic field in Boozer coordinates at $s=0.1$. Bottom: rotational transform.
    }
    \label{fig:nfp4_ncoils1_nonplanar_params}
\end{figure}

We show in \cref{fig:nfp4_ncoils1_nonplanar_params} a 4-field period stellarator with 1 coil per half field-period where the fixed boundary equilibrium has an aspect ratio of $A=6.4$.
The independent coil has a total length of 3.3, a maximum curvature of 28.2, a mean-squared curvature of 29.4, and seven Fourier modes per coil, per coordinate.
The minimum separation between coils is 0.15 and the minimum distance between the coils and the plasma is 0.17.
The rotational transform profile is approximately constant of $\iota \simeq 0.26$ with a mirror ratio $\Delta=0.49$.
The normalized field error on the boundary is at most 0.06\%, leading to a good agreement between the fixed boundary VMEC equilibria and the Poincaré sections, as evidenced in \cref{fig:nfp4_ncoils1_nonplanar_params}.
Such a configuration also possesses a large volume of flux surfaces that extend beyond the last closed surface computed using the single-stage approach.
We also note that, although the contours of magnetic field strength close to the axis for this configuration are nearly vertical, the loss of particles due to outward radial drifts is still substantial as this is not the only criterion for omnigenity.
In fact, as it is evidenced in \cref{fig:nfp4_ncoils1_nonplanar_params}, this configuration does not have a straight magnetic axis at the extrema of the magnetic field on-axis $B$.
To obtain such a configuration, a quasi-isodynamic target could be added to the objective function.

\begin{figure}
    \centering
    \includegraphics[clip,trim=0.2cm 0.2cm 0.2cm 0.2cm,width=.39\textwidth]{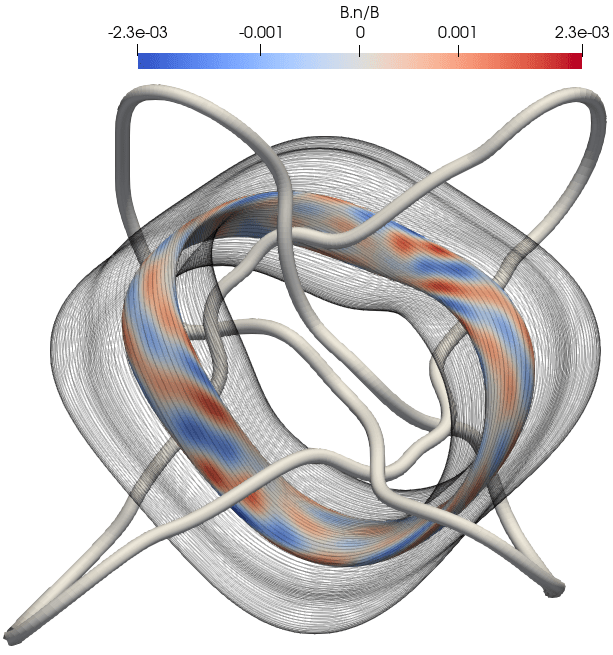}\\
    \includegraphics[clip,trim=0.0cm 0.0cm 0.0cm 0.0cm,width=.33\textwidth]{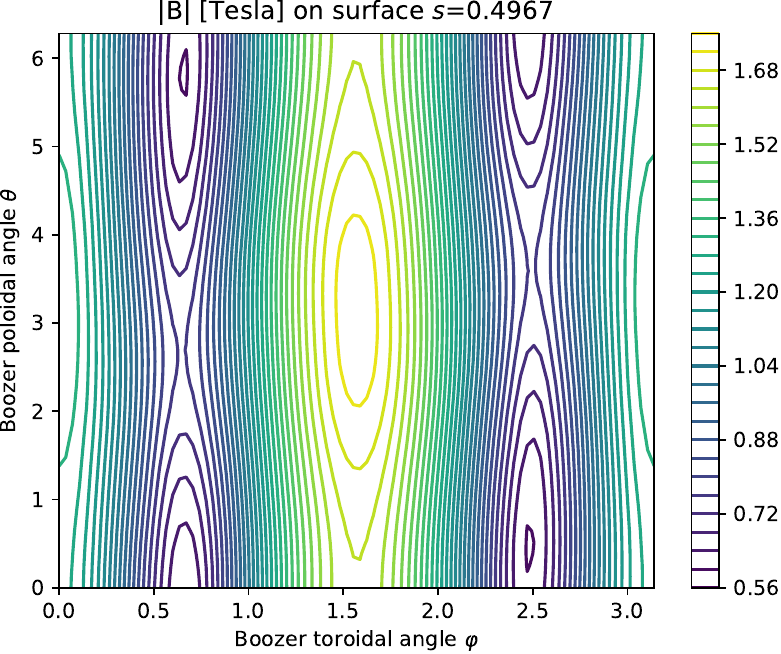}
    \includegraphics[clip,trim=1.0cm 7.2cm 9.5cm 0.4cm,width=.22\textwidth]{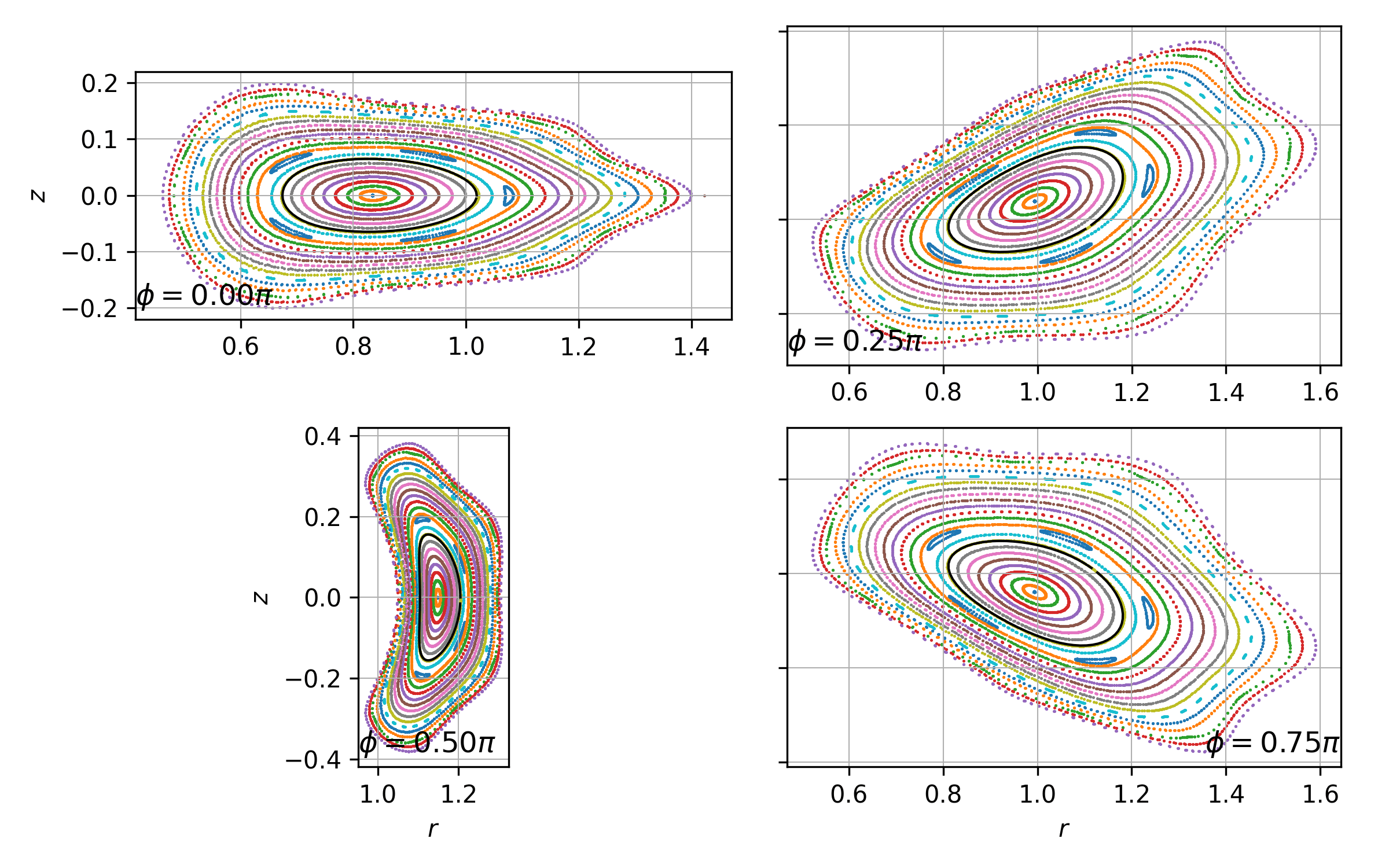}
    \includegraphics[clip,trim=0.1cm 11.9cm 23.0cm 0.7cm,width=.25\textwidth]{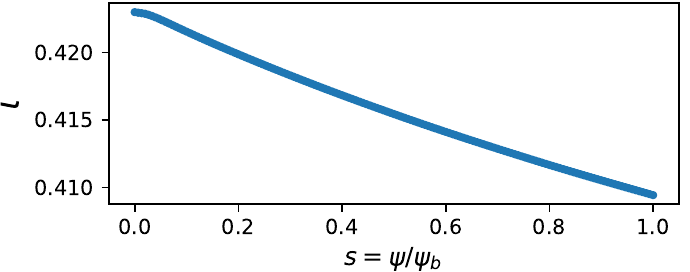}
    \caption{
    Two-field period stellarator with a total of two non-planar coils. Top: three-dimensional view of the fixed boundary magnetic field, the normalized error coil field $\mathbf B \cdot \mathbf n/B$, and magnetic field lines in black. Middle left: Field line tracing at $\phi=0$ and the plasma boundary (in black). Middle right: Contours of constant magnetic field in Boozer coordinates at $s=0.5$. Bottom: rotational transform.
    }
    \label{fig:nfp2_ncoils2_nonplanar_params}
\end{figure}

We show in \cref{fig:nfp2_ncoils2_nonplanar_params} a 2-field period stellarator with a total of 2 coils where the fixed boundary equilibrium has an aspect ratio of $A=8.45$.
Such coils are similar to the ones obtained in Ref. \cite{Kaptanoglu2023} using topology optimization.
Each coil has a total length of 12.0, a maximum curvature of 4.0 and 3.9, a mean-squared curvature of 3.7 and 3.4, respectively, and seven Fourier modes per coil, per coordinate.
The normalized coil currents are 1 and -0.75.
The minimum separation between coils is 0.23 and the minimum distance between the coils and the plasma is 0.17.
The rotational transform profile is approximately constant of $\iota \simeq 0.415$ with a mirror ratio $\Delta=0.43$.
The normalized field error on the boundary is at most $2.2\times 10^{-3}$, leading to an extremely good agreement between the fixed boundary VMEC equilibria and the Poincaré sections, as evidenced in \cref{fig:nfp4_ncoils1_nonplanar_params}.
Such a configuration also possesses a large volume of flux surfaces that extend beyond the last closed surface computed using the single-stage approach.

\section{Quasisymmetric Stellarators with a Low Number of Coils}
\label{sec:qslowcoils}

In this section, we showcase that single-stage optimization targeting quasisymmetry, \cref{eq:fqs}, can still lead to stellarators with a lower number of simplified coils.
We show QA and QH examples at moderate to high aspect ratio from the QUASR database \cite{Giuliani2023a} that have only one and two independent coils, respectively, that are far away from the plasma and small curvature.
Then, we show a QH example using the single-stage fixed boundary method \cite{Jorge2023} at a smaller aspect ratio but with three independent coils and a smaller plasma-to-coil distance.
All reported root of quasisymmetry errors measured below correspond to $\sqrt{f_{\text{QS}}}$ defined in \cref{eq:fqs}.

\subsection{Quasi-Axisymmetric Stellarator}

\begin{figure}
    \centering
    \includegraphics[clip,trim=0.2cm 0.2cm 0.4cm 1.5cm,width=.41\textwidth]{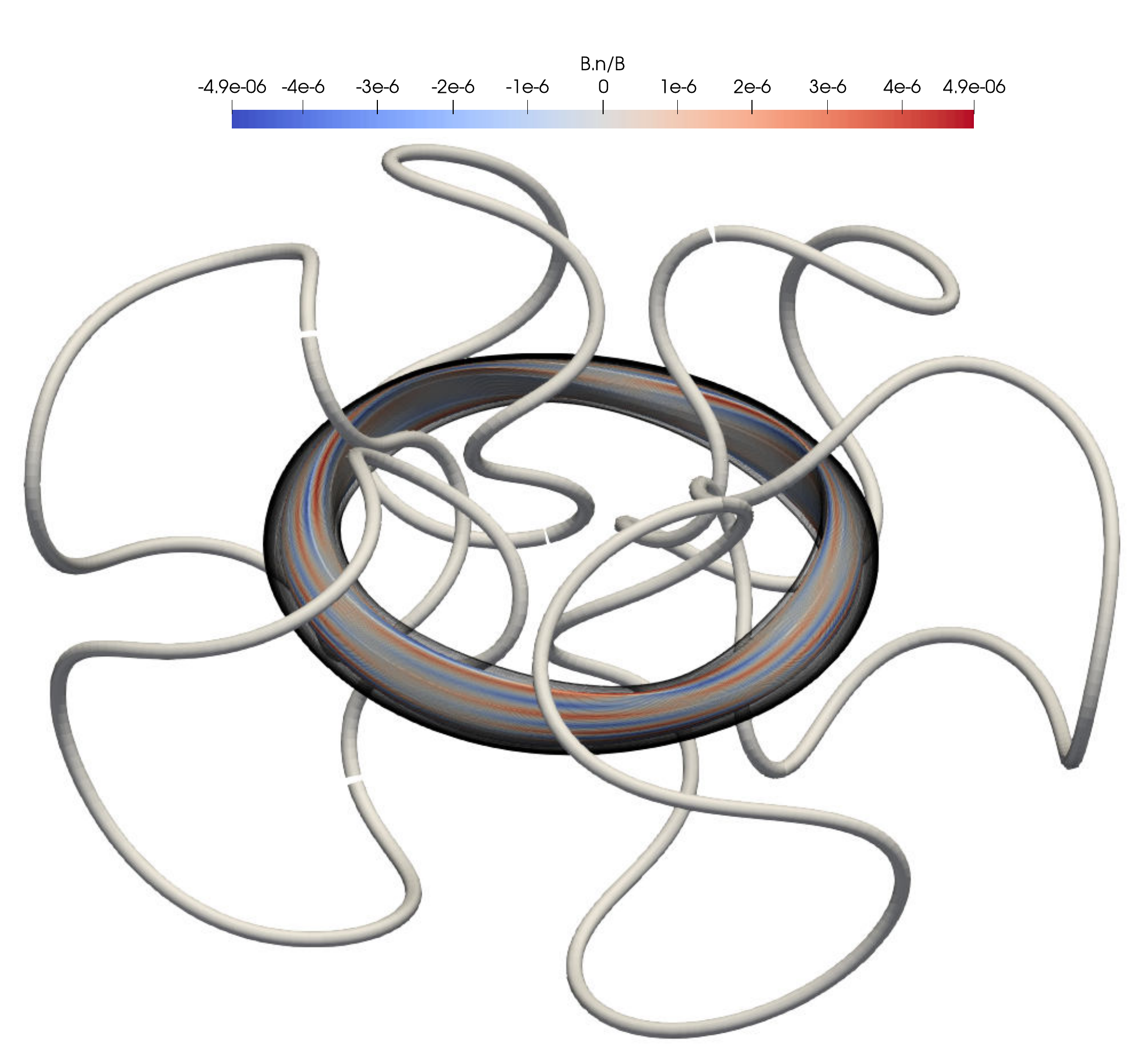}\\
    \includegraphics[clip,trim=0.0cm 0.0cm 0.0cm 0.0cm,width=.38\textwidth]{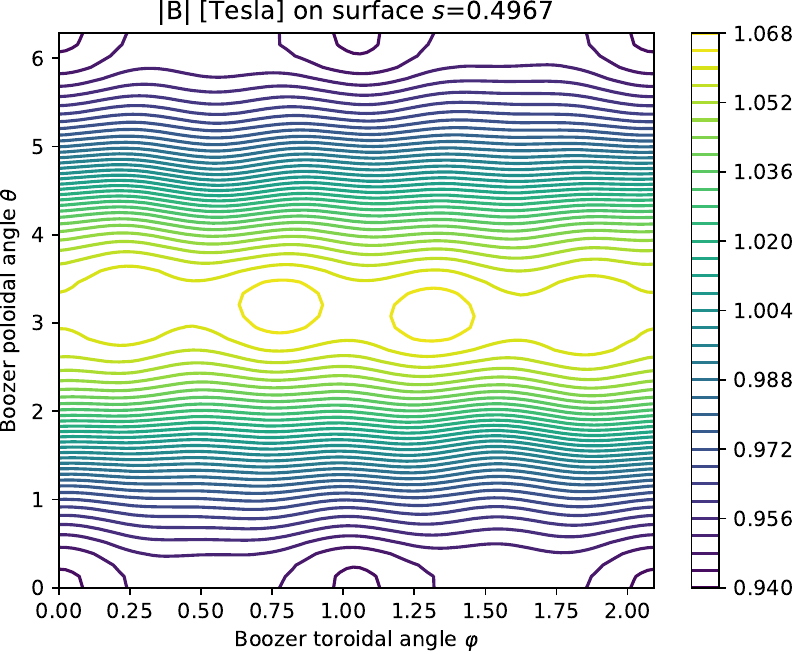}
    \includegraphics[clip,trim=1.3cm 6.6cm 9.2cm 0.2cm,width=.19\textwidth]{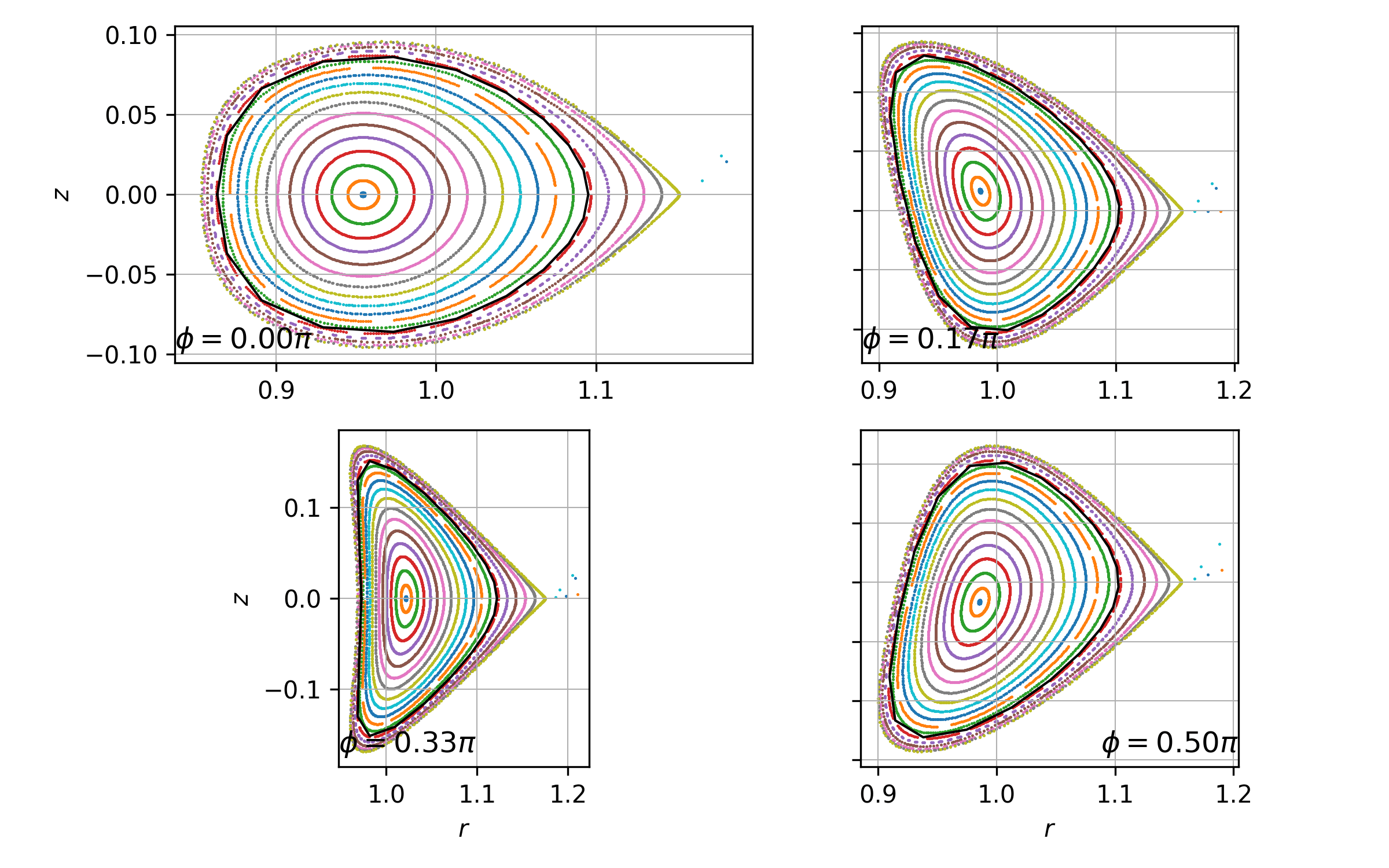}
    \includegraphics[clip,trim=0.1cm 11.9cm 22.9cm 0.7cm,width=.28\textwidth]{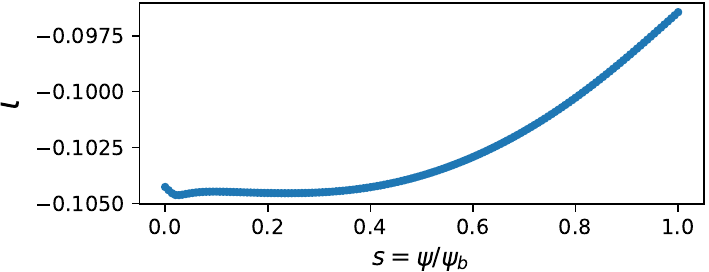}
    \caption{
    Two-field period quasi-axisymmetric stellarator with one non-planar coil per half field-period (QUASR device ID: 0635650). Top: three-dimensional view of the fixed boundary magnetic field, the normalized error coil field $\mathbf B \cdot \mathbf n/B$, and magnetic field lines in black. Middle: Contours of constant magnetic field in Boozer coordinates at $s=0.5$. Bottom left: Field line tracing at $\phi=0$ and the plasma boundary (in black). Bottom right: rotational transform.
    }
    \label{fig:QUASR_QA_ncoils1_params}
\end{figure}

We show in \cref{fig:QUASR_QA_ncoils1_params} QUASR device ID: 0635650, a 2-field period quasi-axisymmetric stellarator with 1 coil per half field-period where the fixed boundary equilibrium has an aspect ratio of $A=10.0$.
The independent coil has a total length of 7.3, a maximum curvature of 3.8, a mean-squared curvature of 5.0, and sixteen Fourier modes per coil, per coordinate.
The minimum separation between coils is 0.12 and the minimum distance between the coils and the plasma is 0.32.
The rotational transform profile is approximately constant of $\iota \simeq 0.1$ with a mirror ratio $\Delta=0.006$.
The normalized field error on the boundary is at most $4.9\times10^{-6}$, leading to an extremely good agreement between the fixed boundary VMEC equilibria and the Poincaré sections, as evidenced in \cref{fig:QUASR_QA_ncoils1_params}.
However, due to the fact that this configuration has a low rotational transform and a root of the quasisymmetry error of $3.0\times10^{-3}$, the loss fraction of 3500 alpha particles launched from $s=0.25$ at $0.01$s is approximately 44\%.

\subsection{Quasi-Helically Symmetric Stellarator}

\begin{figure}
    \centering
    \includegraphics[clip,trim=1.5cm 2.9cm 2.4cm 0.5cm,width=.41\textwidth]{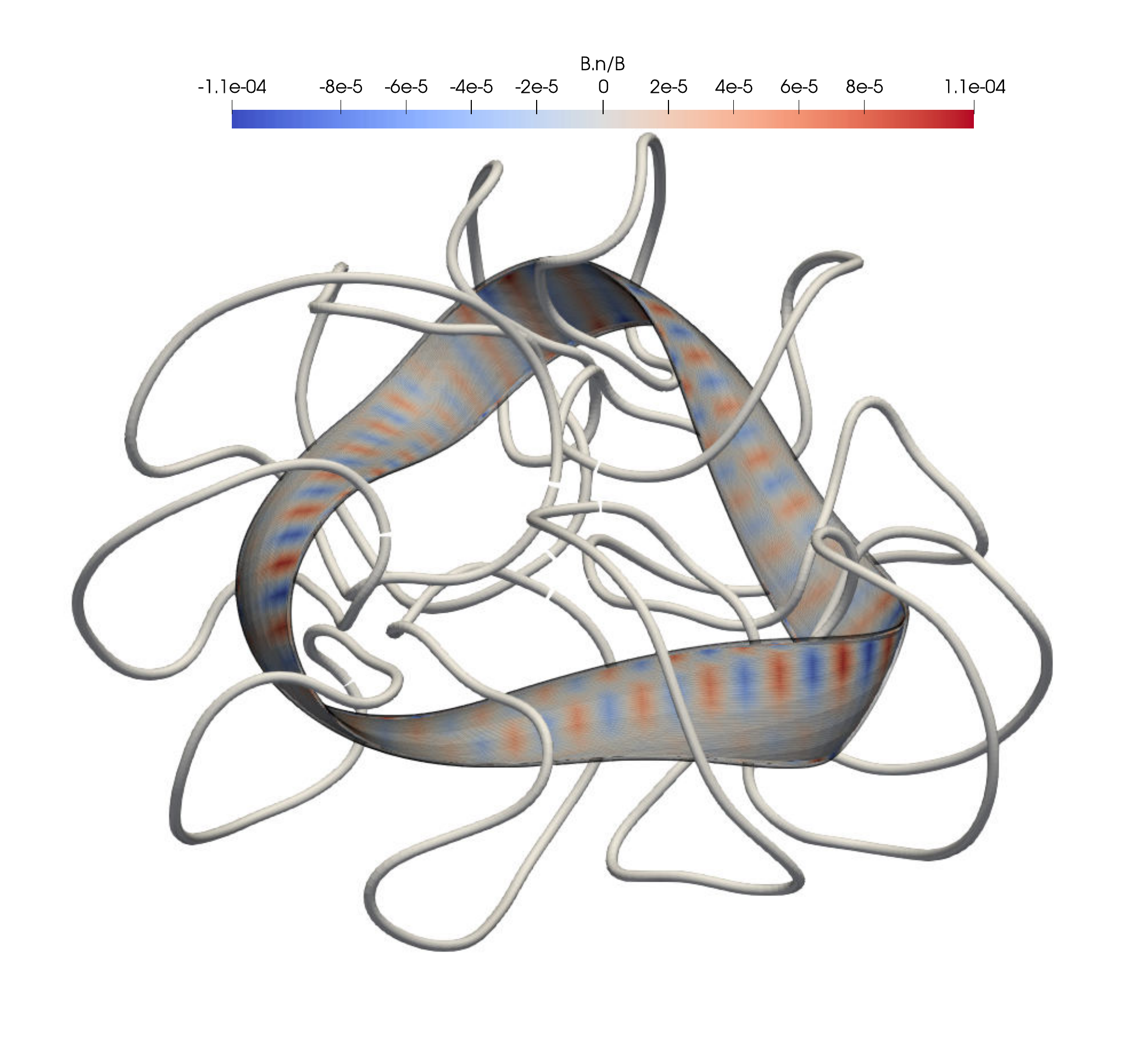}\\
    \includegraphics[clip,trim=5.1cm 6.3cm 13.1cm 0.3cm,width=.10\textwidth]{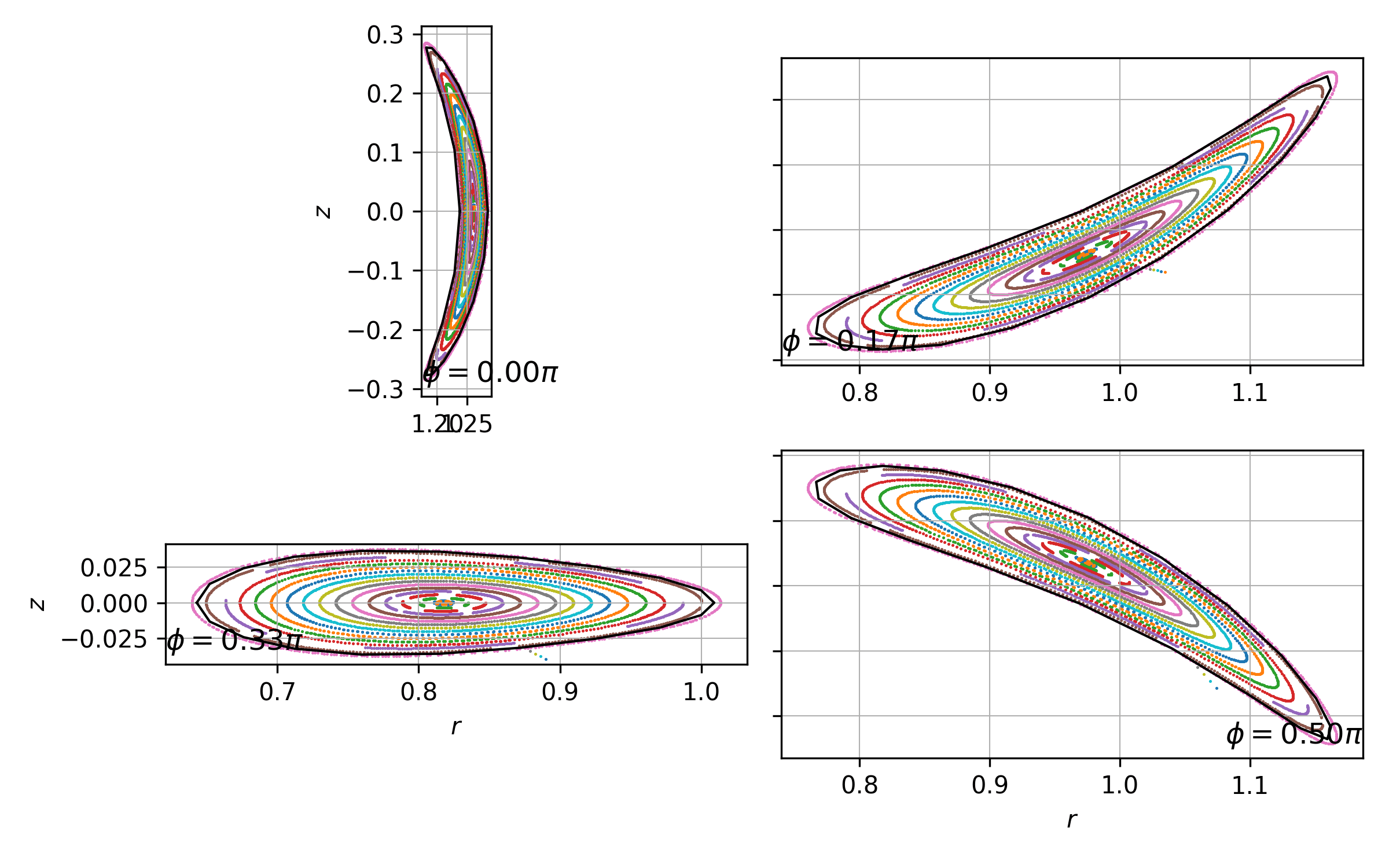}
    \includegraphics[clip,trim=0.0cm 0.0cm 0.0cm 0.0cm,width=.37\textwidth]{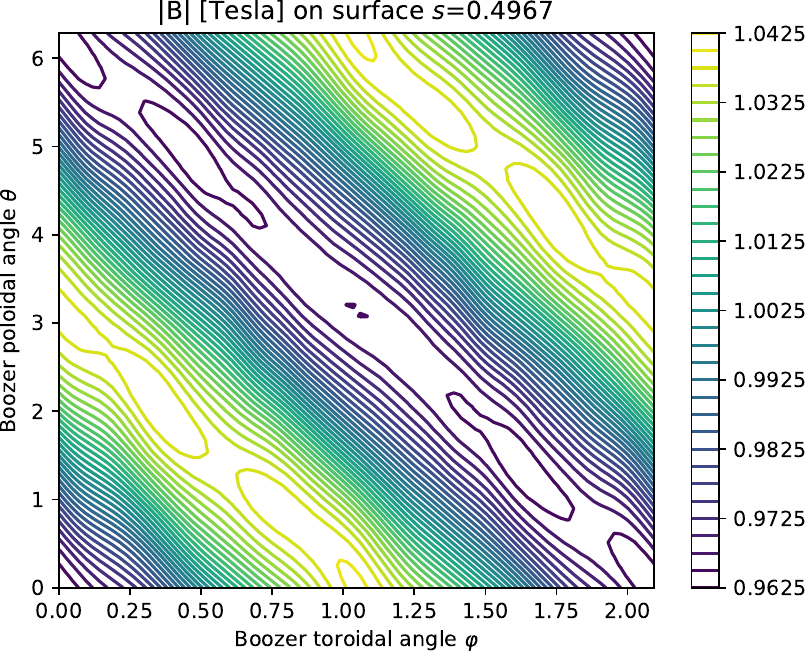}
    \includegraphics[clip,trim=0.1cm 11.9cm 22.6cm 0.7cm,width=.35\textwidth]{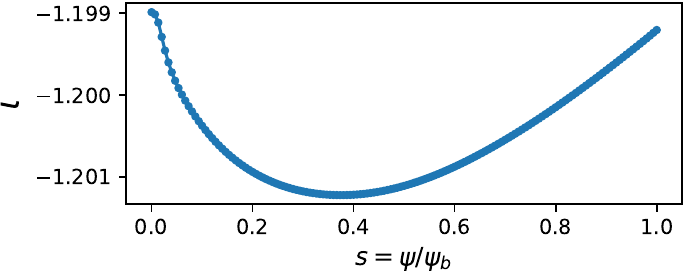}
    \caption{
    Three-field period quasi-helically symmetric stellarator with two non-planar coils per half field-period (QUASR device ID: 1426796). Top: three-dimensional view of the fixed boundary magnetic field, the normalized error coil field $\mathbf B \cdot \mathbf n/B$, and magnetic field lines in black. Middle left: Field line tracing at $\phi=0$ and the plasma boundary (in black). Middle right: Contours of constant magnetic field in Boozer coordinates at $s=0.5$. Bottom: rotational transform.
    }
    \label{fig:QUASR_QH_ncoils2_params}
\end{figure}

We show in \cref{fig:QUASR_QH_ncoils2_params} QUASR device ID: 1426796, a 3-field period quasi-helically symmetric stellarator with 2 coils per half field-period. where the fixed boundary equilibrium has an aspect ratio of $A=12.0$
The independent coils have a total length of 4.0 and 5.2 each, a maximum curvature of 5.3 and 6.1, a mean-squared curvature of 9.6 and 9.0, and sixteen Fourier modes per coil, per coordinate.
The minimum separation between coils is 0.10 and the minimum distance between the coils and the plasma is 0.20.
The rotational transform profile is approximately constant of $\iota \simeq 1.2$ with a mirror ratio $\Delta=0.004$.
The normalized field error on the boundary is at most $1.1\times10^{-4}$, leading to an extremely good agreement between the fixed boundary VMEC equilibria and the Poincaré sections, as evidenced in \cref{fig:QUASR_QH_ncoils2_params}.
Due to the large rotational transform and a root of the quasisymmetry error of $7.0\times10^{-2}$, the loss fraction of 3500 alpha particles launched from $s=0.25$ at $0.01$s is approximately 0.12\%.

\begin{figure}
    \centering
    \includegraphics[clip,trim=3.2cm 0.0cm 2.4cm 0.4cm,width=.41\textwidth]{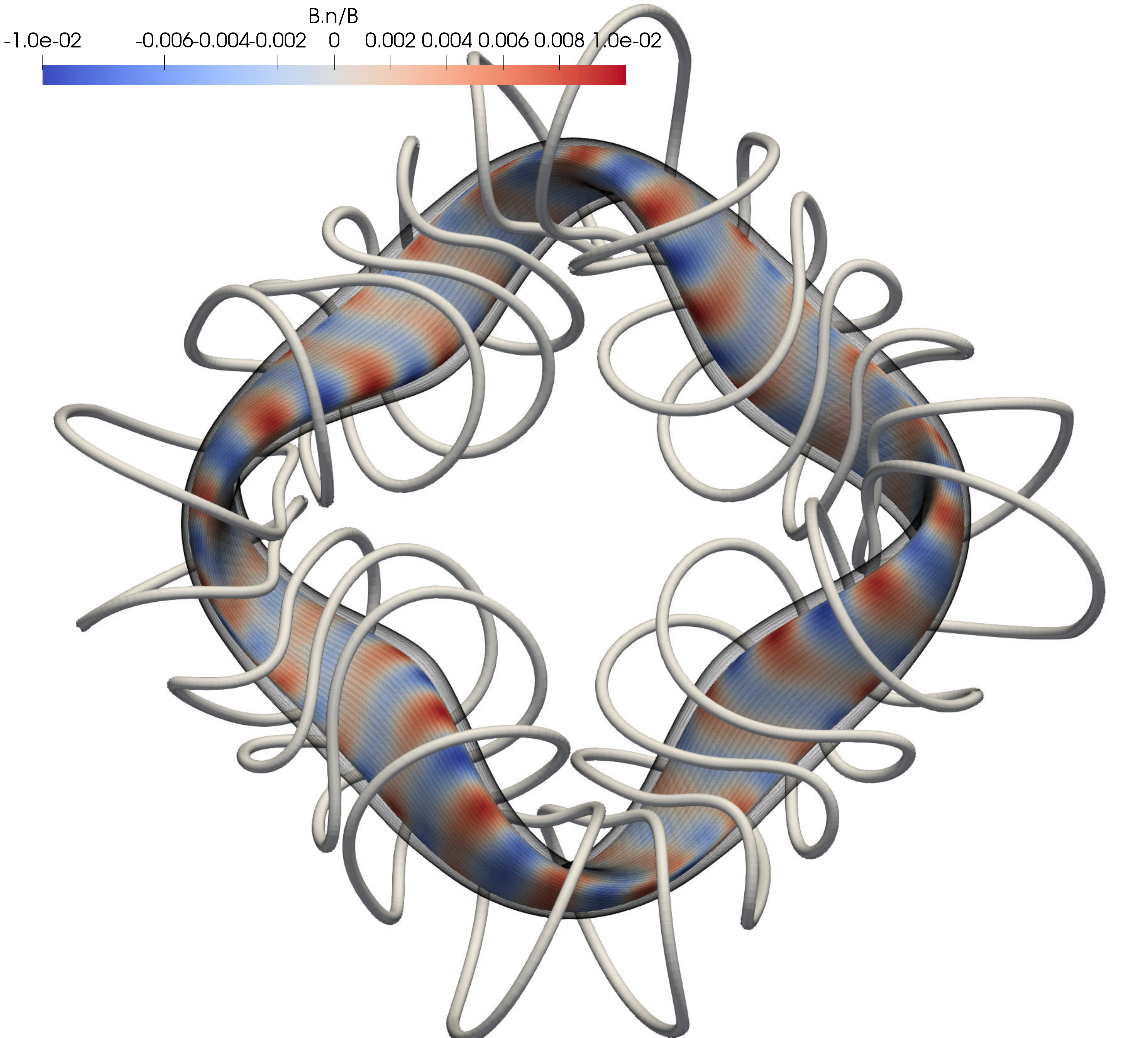}\\
    \includegraphics[clip,trim=4.7cm 6.3cm 12.9cm 0.3cm,width=.125\textwidth]{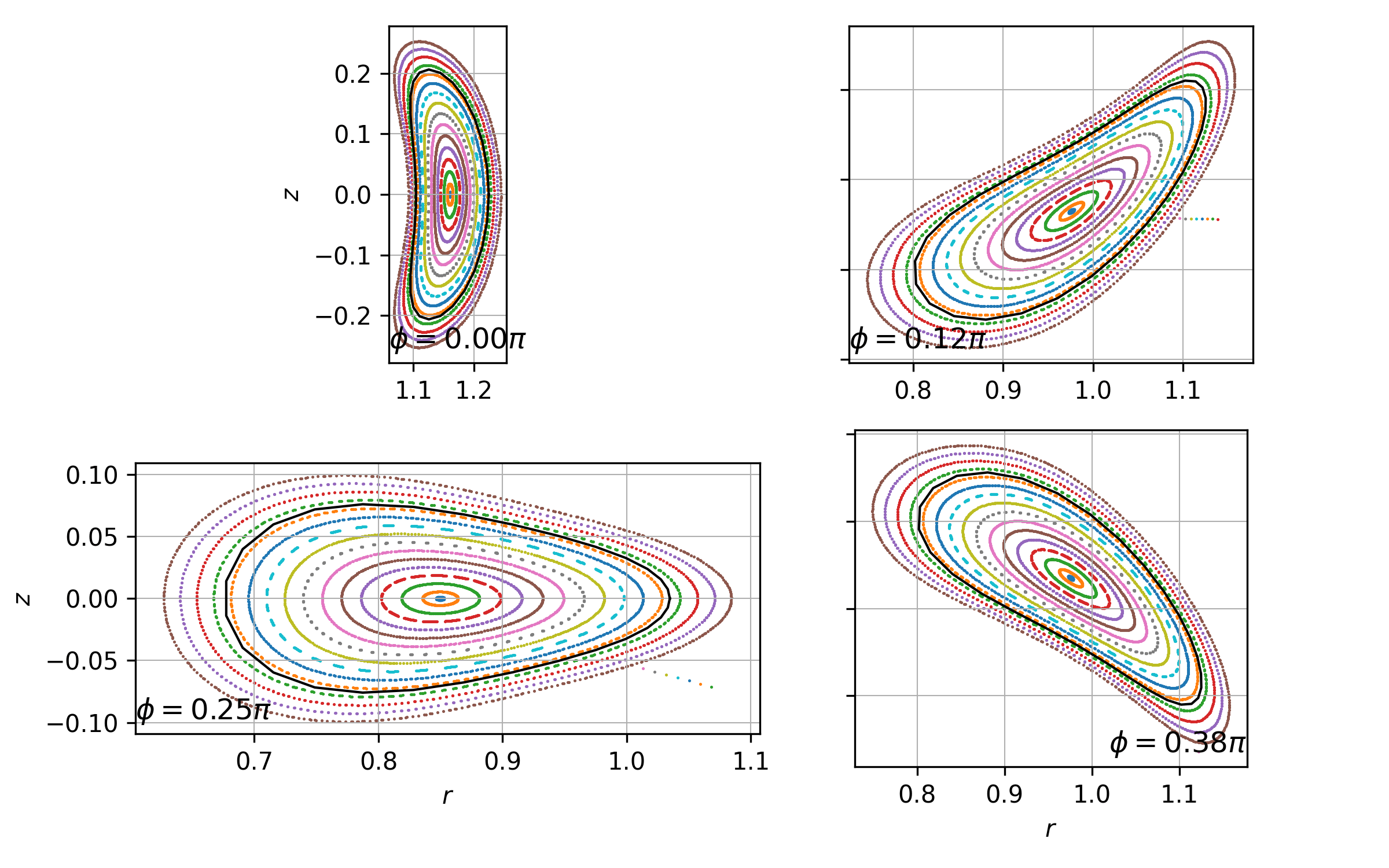}
    \includegraphics[clip,trim=0.0cm 0.0cm 0.0cm 0.0cm,width=.35\textwidth]{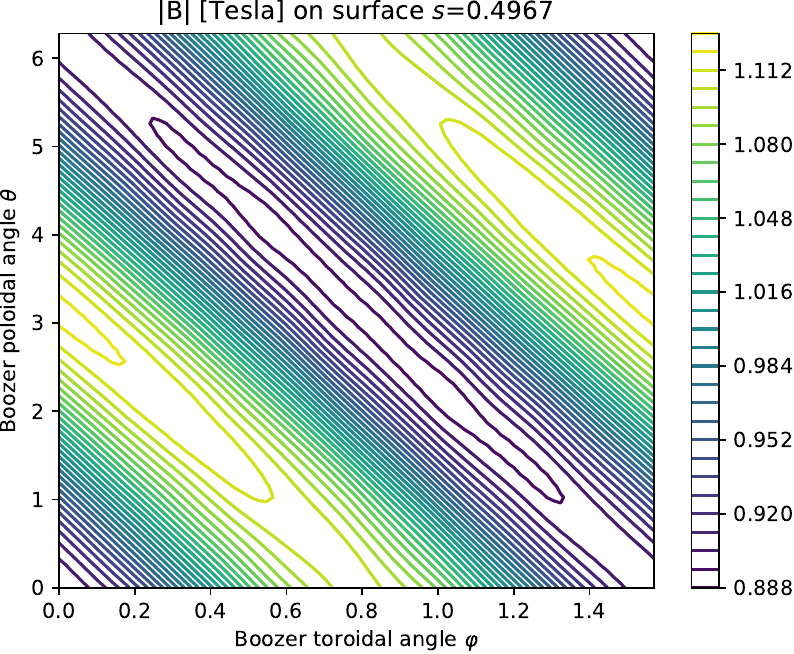}
    \includegraphics[clip,trim=0.1cm 11.9cm 22.9cm 0.7cm,width=.35\textwidth]{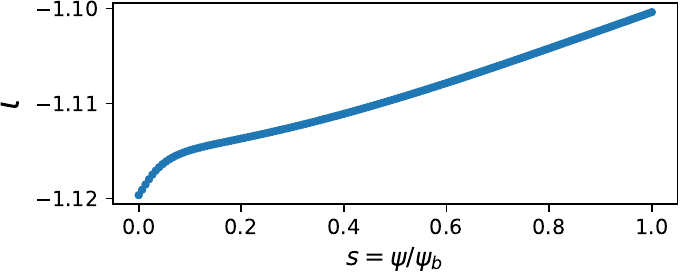}
    \caption{
    Four-field period quasi-helically symmetric stellarator with three non-planar coils per half field-period. Top: three-dimensional view of the fixed boundary magnetic field, the normalized error coil field $\mathbf B \cdot \mathbf n/B$, and magnetic field lines in black. Middle left: Field line tracing at $\phi=0$ and the plasma boundary (in black). Middle right: Contours of constant magnetic field in Boozer coordinates at $s=0.5$. Bottom: rotational transform.
    }
    \label{fig:QH_ncoils3_nonplanar_params}
\end{figure}

We show in \cref{fig:QH_ncoils3_nonplanar_params} a 4-field period quasi-helically symmetric stellarator with 3 coils per half field-period, obtained using the single-stage approach via fixed-boundary optimization that has an aspect ratio of $A=8.5$.
For comparison, the vacuum quasi-helical symmetric coil configuration of Ref. \!\![\onlinecite{Wiedman2023}] with aspect ratio $A=8$ has 5 coils per half field-period and when scaled from its ARIES-CS minor radius of $1.7$m to the minor radius of $1/A=0.118$m by a factor of $0.118/1.7\simeq7\times 10^{-2}$, have an average length of 2.5, coil to coil distance of 0.076, a distance between the surface and the coils of 0.11, a curvature between 8.6 and 11, and a mean-squared curvature of 1.
The 3 independent coils have a total length of 2.7 each, a maximum curvature of 10.2, 7.8, and 7.4, a mean-squared curvature of 10.2, 11.3, and 12.5, and five Fourier modes per coil, per coordinate.
The minimum separation between coils is 0.08 and the minimum distance between the coils and the plasma is 0.15.
The rotational transform profile is approximately constant of $\iota \simeq 1.11$ with a mirror ratio $\Delta=0.014$.
The normalized field error on the boundary is at most 1\%, leading to a good agreement between the fixed boundary VMEC equilibria and the Poincaré sections, as evidenced in \cref{fig:QH_ncoils3_nonplanar_params}.
Due to the large rotational transform and a total quasisymmetry residual of $5.3\times10^{-3}$ using \cref{eq:fqs}, the loss fraction of 3500 alpha particles launched from $s=0.25$ at $0.01$s is 0.
Furthermore, it is seen that there is still an appreciable volume of flux surfaces extending beyond the fixed boundary solution.

\section{\label{sec:QAtrimcoil} Quasi-Axisymmetric Stellarators with External Coils}

External non-axisymmetric coils, such as trim or error field correction coils are used, for example, to induce Resonant Magnetic Perturbations for Edge Localized Mode Control (see, for example, Ref. \!\![\onlinecite{Yang2020}]).
At the W7-X device, external trim coils are used to fine-tune the main magnetic field during plasma operation \cite{Risse2011}.
However, such studies are usually done as an extra step once the magnetic field equilibrium and set of coils is already established.
In this section, we show how single-stage optimization can be used to incorporate external coils, i.e., coils that are not linked with the plasma, in the total magnetic field used to produce the plasma.
They introduce an additional radial magnetic field that helps shape the plasma and contribute to the rotational transform.
Such coils are initialized away from the plasma, with their centers at $y=[-1.2, 1.3]$ and $z=[-0.85, -0.95]$, and a radius of 0.6.
Their parametrization follows the space curve description of \cref{sec:coils}.

\subsection{One Coil per half field-period}

\begin{figure}
    \centering
    \includegraphics[clip,trim=0.2cm 8.0cm 5.4cm 2.4cm,width=.41\textwidth]{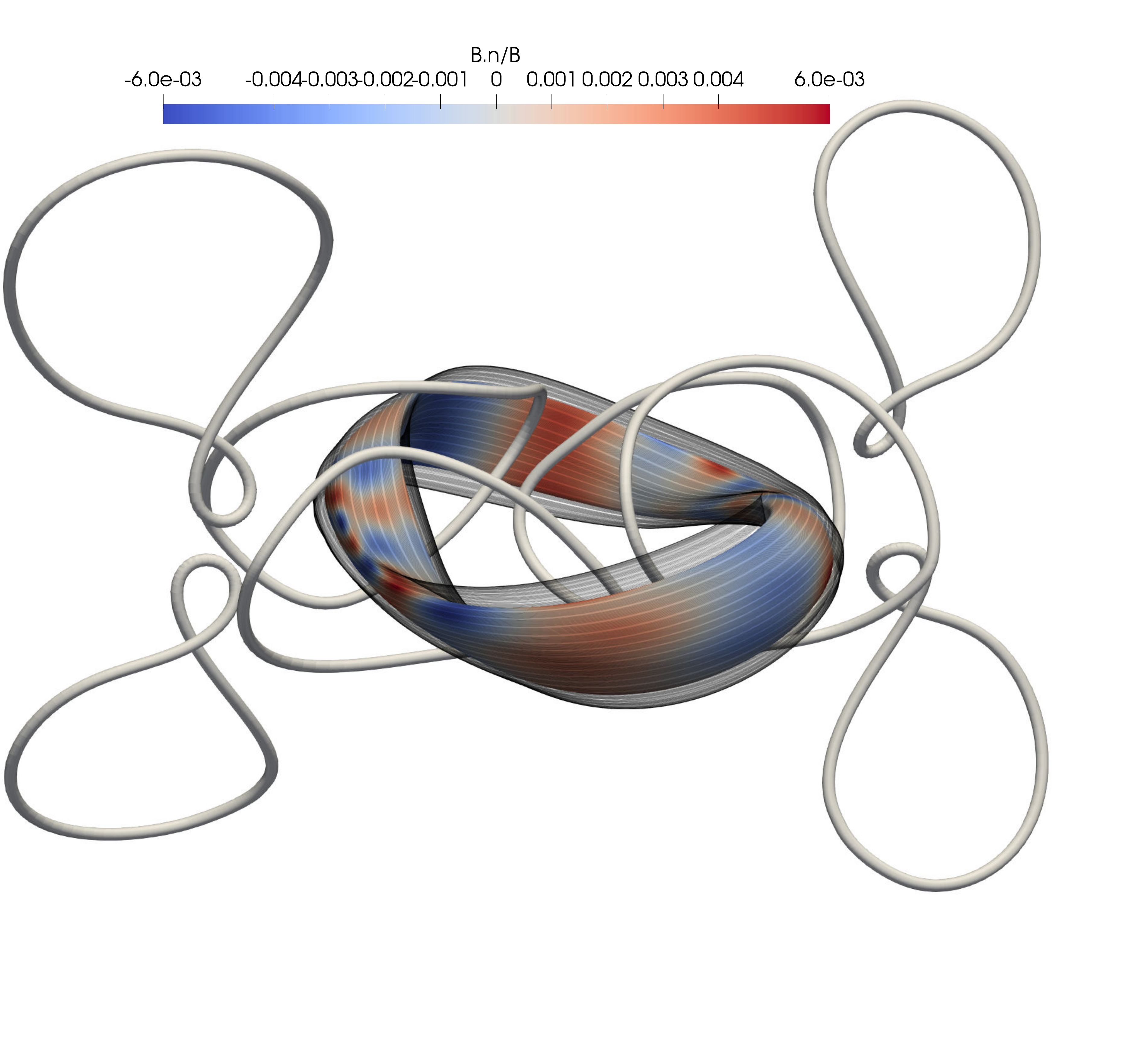}\\
    \includegraphics[clip,trim=4.95cm 0.8cm 12.5cm 6.2cm,width=.135\textwidth]{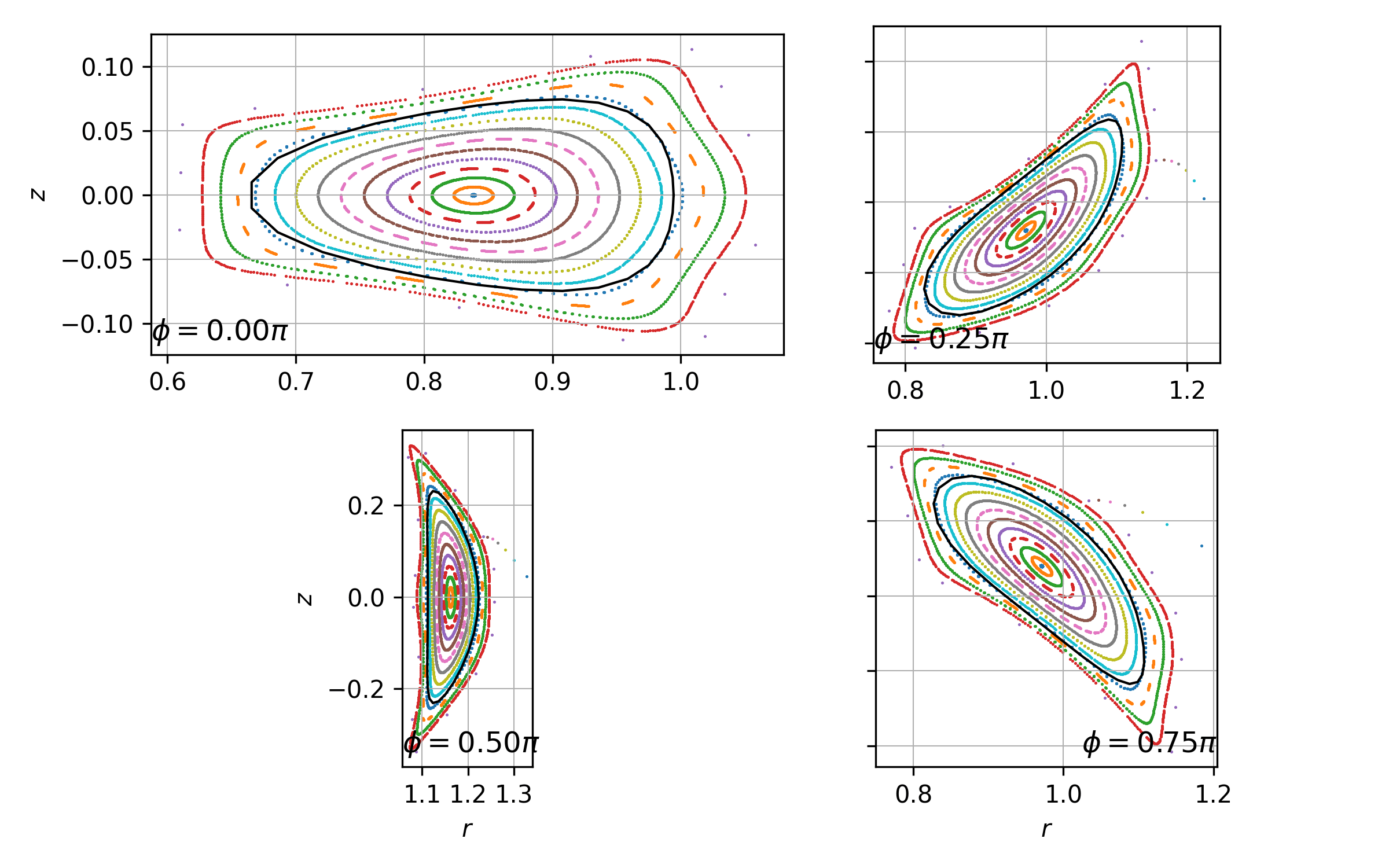}
    \includegraphics[clip,trim=0.0cm 0.0cm 0.0cm 0.0cm,width=.34\textwidth]{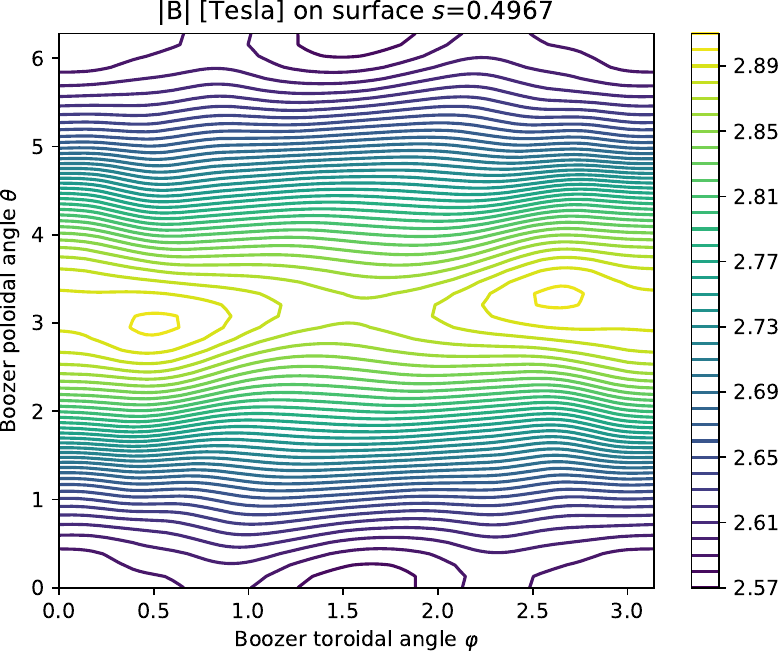}
    \includegraphics[clip,trim=0.1cm 11.9cm 22.9cm 0.7cm,width=.35\textwidth]{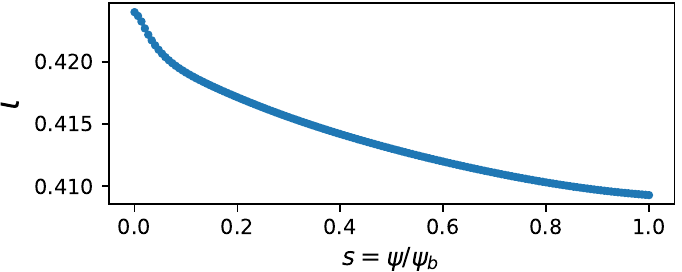}
    \caption{
    Two-field period quasi-axisymmetric stellarator with one non-planar modular coil and one external trim coil per half field-period. Top: three-dimensional view of the fixed boundary magnetic field, the normalized error coil field $\mathbf B \cdot \mathbf n/B$, and magnetic field lines in black. Upper middle: Contours of constant magnetic field in Boozer coordinates at $s=0.5$. Lower middle: Field line tracing at $\phi=0.5\pi$ and the plasma boundary (in black). Bottom: rotational transform.
    }
    \label{fig:QA_ncoils1_extra_params}
\end{figure}

We show in \cref{fig:QA_ncoils1_extra_params} a 2-field period quasi-axisymmetric stellarator with 1 modular coil per half field-period and 1 external trim coil where the fixed boundary equilibrium has an aspect ratio of $A=8.9$.
The 2 independent coils have a total length of 4.5 each, a maximum curvature of 9.0, a mean-squared curvature of 4.0 and 9.1, and five Fourier modes per coil, per coordinate.
The minimum separation between coils is 0.16 and the minimum distance between the coils and the plasma is 0.43.
The rotational transform profile is approximately constant of $\iota \simeq 0.415$ with a mirror ratio $\Delta=0.0014$.
The normalized field error on the boundary is at most 0.6\%, leading to a good agreement between the fixed boundary VMEC equilibria and the Poincaré sections, as evidenced in \cref{fig:QA_ncoils1_extra_params}.
Due to the target rotational transform of 0.41 and a total quasisymmetry residual of $1.4\times10^{-3}$ using \cref{eq:fqs}, the loss fraction of 3500 alpha particles launched from $s=0.25$ at $0.01$s is 14\%.
The current of the modular coil is fixed at $10^5$A during the optimization, leading to a current of $1.1\times 10^6$A for the external trim coil.

\subsection{Two Coils per half field-period}

\begin{figure}
    \centering
    \includegraphics[clip,trim=0.2cm 4.8cm 0.2cm 3.7cm,width=.45\textwidth]{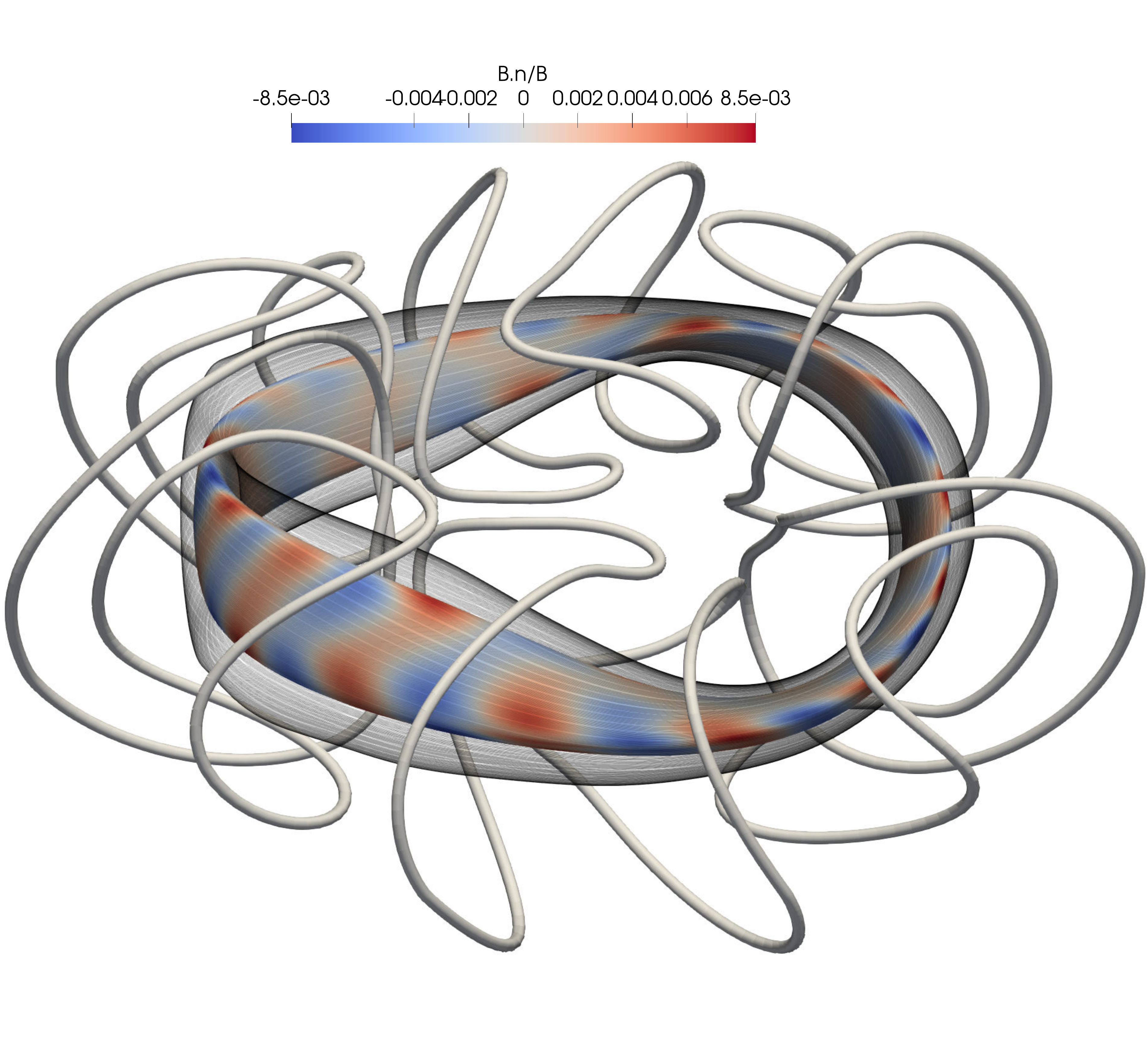}\\
    \includegraphics[clip,trim=4.95cm 0.8cm 12.9cm 6.2cm,width=.125\textwidth]{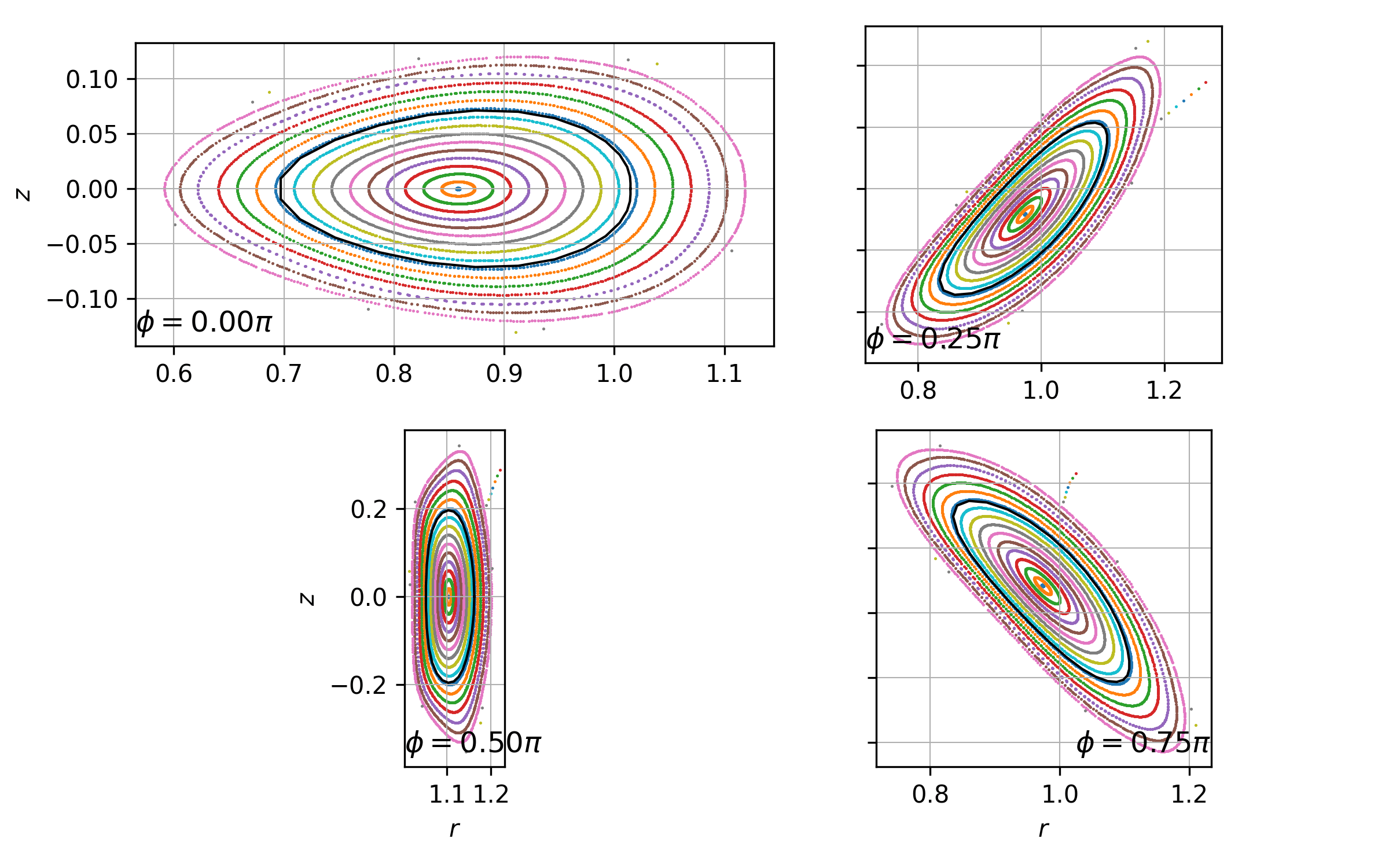}
    \includegraphics[clip,trim=0.0cm 0.0cm 0.0cm 0.0cm,width=.35\textwidth]{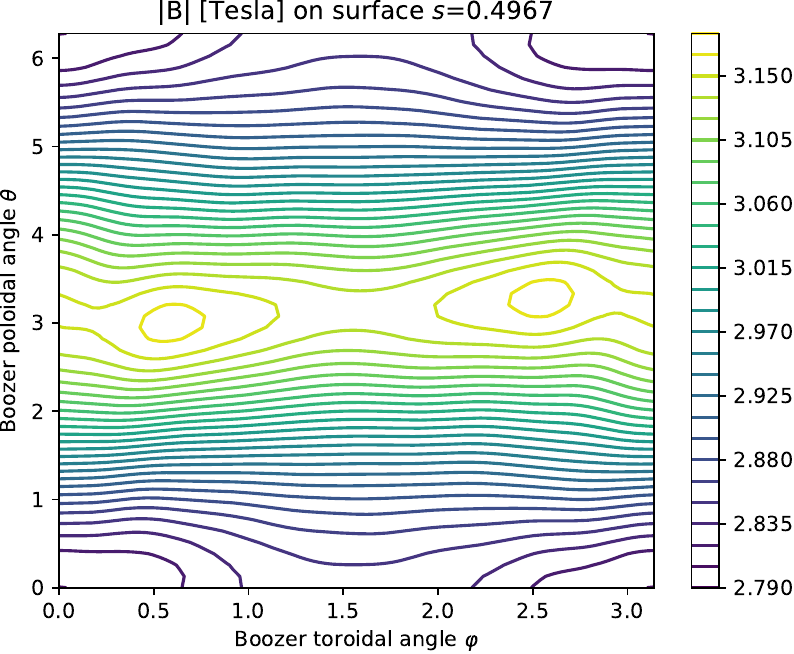}
    \includegraphics[clip,trim=0.1cm 11.9cm 22.8cm 0.7cm,width=.35\textwidth]{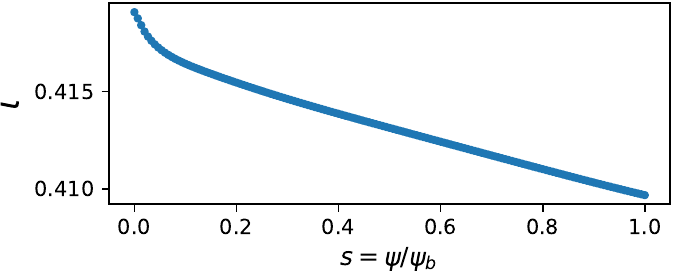}
    \caption{
    Two-field period quasi-axisymmetric stellarator with two non-planar modular coils and one external trim coil per half field-period. Top: three-dimensional view of the fixed boundary magnetic field, the normalized error coil field $\mathbf B \cdot \mathbf n/B$, and magnetic field lines in black. Upper middle: Contours of constant magnetic field in Boozer coordinates at $s=0.5$. Lower middle: Field line tracing at $\phi=0.5\pi$ and the plasma boundary (in black). Bottom: rotational transform.
    }
    \label{fig:QA_ncoils2_extra_params}
\end{figure}

We show in \cref{fig:QA_ncoils2_extra_params} a 2-field period quasi-axisymmetric stellarator with 2 modular coils per half field-period and 1 external trim coil where the fixed boundary equilibrium has an aspect ratio of $A=9.3$.
The 3 independent coils have a total length of 4.2 each, a maximum curvature of 9.0 for the modular coils and 8.3 for the trim coil, a mean-squared curvature of 9.0 and 4.4 for the modular coils and 9.0 for the trim coil, and five Fourier modes per coil, per coordinate.
The minimum separation between coils is 0.20 and the minimum distance between the coils and the plasma is 0.25.
The rotational transform profile is approximately constant of $\iota \simeq 0.41$ with a mirror ratio $\Delta=0.0058$.
The normalized field error on the boundary is at most 0.85\%, leading to a good agreement between the fixed boundary VMEC equilibria and the Poincaré sections, as evidenced in \cref{fig:QA_ncoils2_extra_params}.
Due to the target rotational transform of 0.41 and a total quasisymmetry residual of $1.2\times10^{-3}$ using \cref{eq:fqs}, the loss fraction of 3500 alpha particles launched from $s=0.25$ at $0.01$s is 12\%.
The current of the first modular coil is fixed at $10^5$A during the optimization, leading to a current of $2.05\times 10^5$A for the second modular coil and $5.3\times 10^4$A for the external trim coil.
We also find that there is a large volume of nested flux surfaces outside the boundary computed by VMEC.

\section{Quasisymmetry with Helical Coils}
\label{sec:helicalsection}

Due to the intrinsically helical nature of stellarators, another option for coils is the use of a helical coil \cite{Boozer2024}.
The advantages of coils are well documented \cite{Yamaguchi2019}, such as relatively open access to the plasma, smaller coil ripple, and good particle confinement.
In the following, we consider two different types of helical coils: helical coils that wind around a circular torus, leading to a positive magnetic well stellarator, and helical coils that are not bound to such a winding surface, leading to better confinement properties.

\subsection{Two Helical Coils in a Circular Torus and Positive Magnetic Well}
\label{sec:helicalcircular}

\begin{figure}
    \centering
    \includegraphics[clip,trim=1.3cm 2.8cm 2.0cm 2.3cm,width=.42\textwidth]{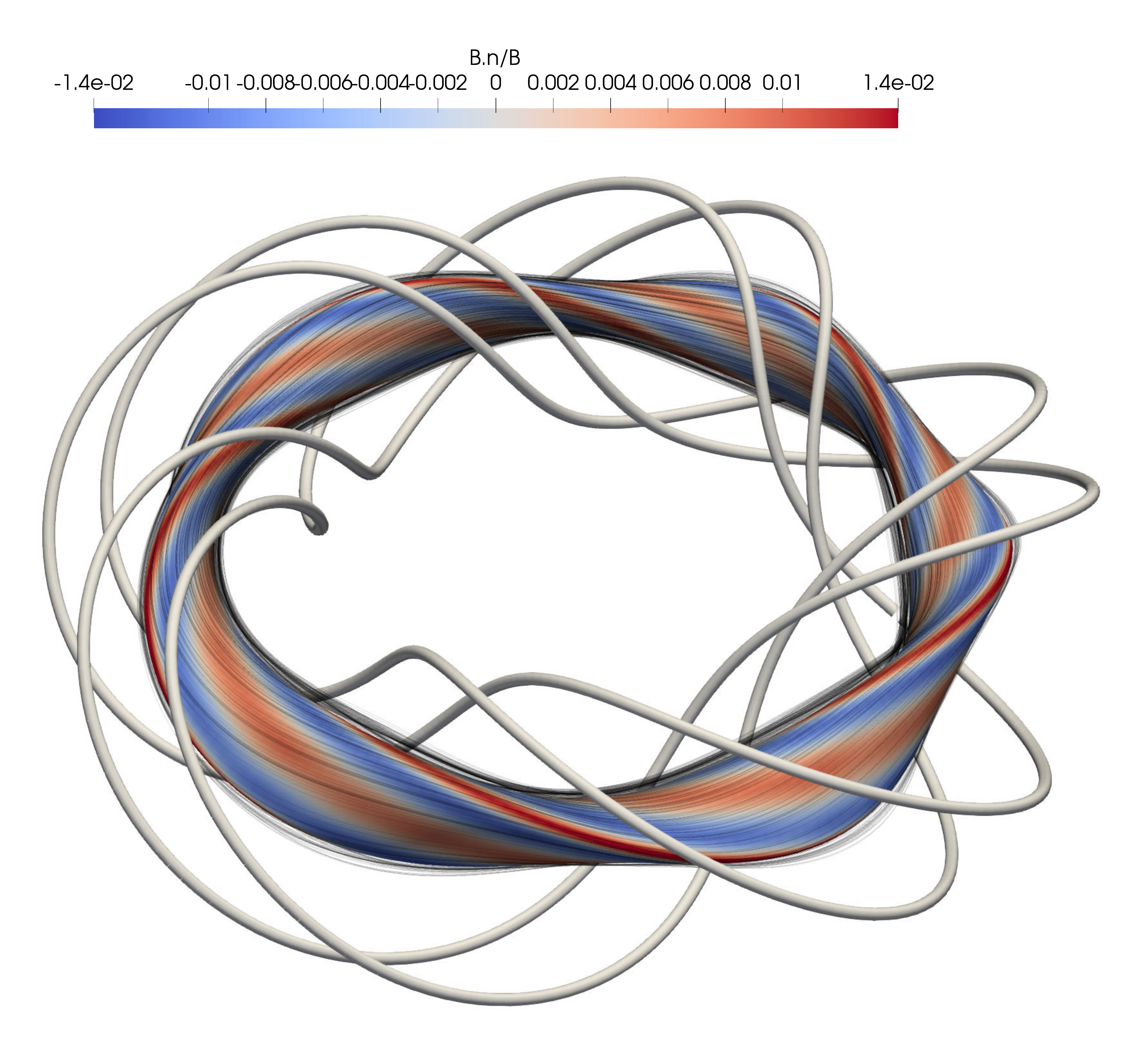}\\
    \includegraphics[clip,trim=4.5cm 0.8cm 12.2cm 6.2cm,width=.16\textwidth]{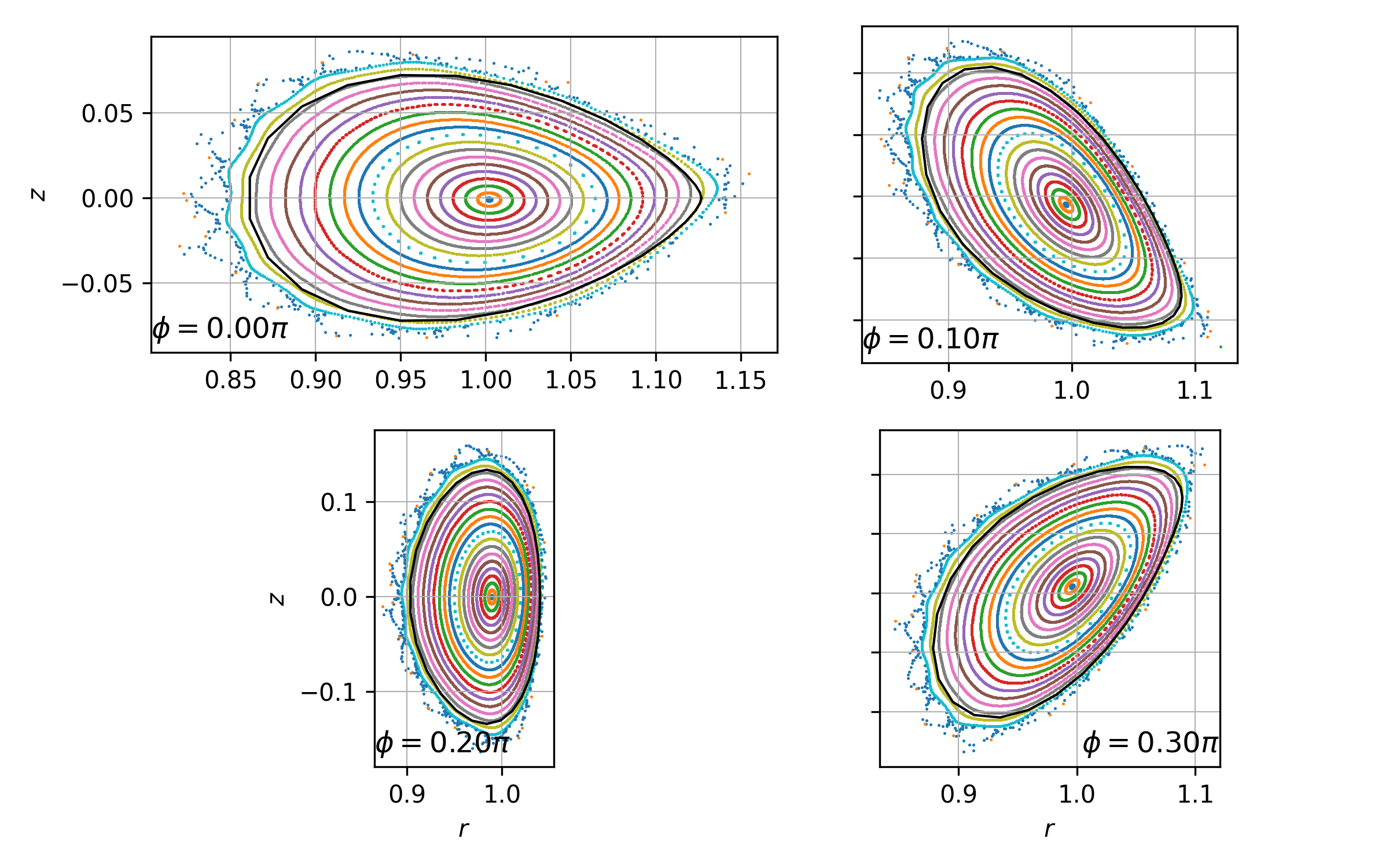}
    \includegraphics[clip,trim=0.0cm 0.0cm 0.0cm 0.0cm,width=.31\textwidth]{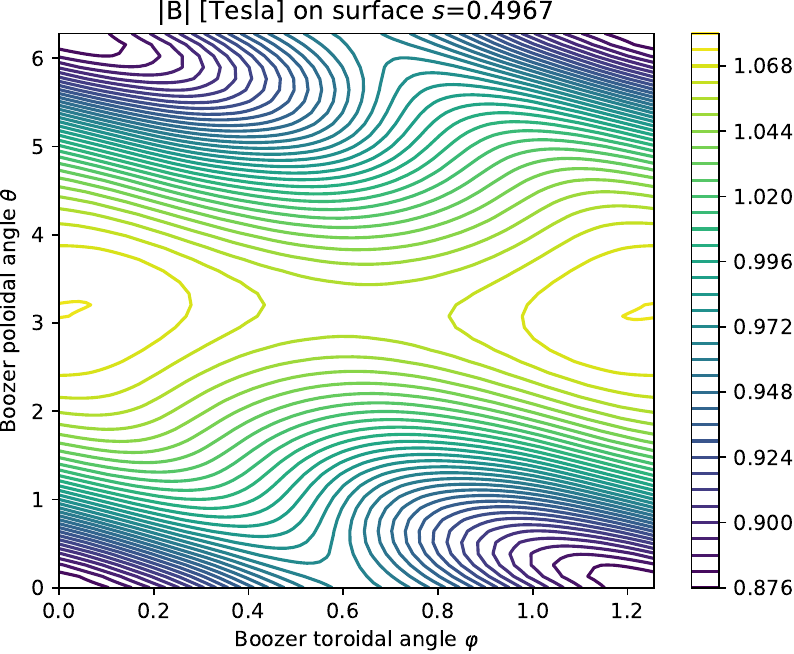}\\
    \includegraphics[clip,trim=0.1cm 11.9cm 22.8cm 0.7cm,width=.36\textwidth]{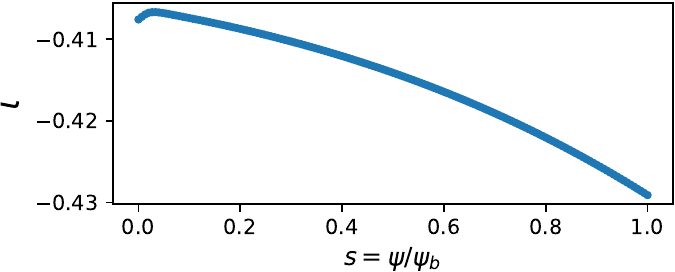}
    \caption{
    Two-field period quasi-axisymmetric stellarator with two helical coils. Top: three-dimensional view of the fixed boundary magnetic field, the normalized error coil field $\mathbf B \cdot \mathbf n/B$, and magnetic field lines in black. Upper middle: Contours of constant magnetic field in Boozer coordinates at $s=0.5$. Lower middle: Field line tracing at $\phi=0.2\pi$ and the plasma boundary (in black). Bottom: rotational transform.
    }
    \label{fig:QA_helical_ncoils2_params}
\end{figure}

We show in \cref{fig:QA_helical_ncoils2_params} a 2-field period quasi-axisymmetric stellarator with two $l_0=5$ helical coils where the fixed boundary equilibrium has an aspect ratio of $A=10.2$.
The field is created using a single helical coil and no additional toroidal magnetic field.
A toroidal magnetic field $\mathbf B = B_0 R_0/R \mathbf e_\phi$ with $(R,\phi,Z)$ the standard cylindrical toroidal angles were added to mimic an additional $1/R$ dependence, with $B_0=1$ T and $R_0=1$ m.
The two independent coils have a total length of 16.8 and 16.7 each, a maximum curvature of 1.7 and 1.6, a mean-squared curvature of 2.3 and 2.2, and seven Fourier modes per coil, per coordinate.
The minimum separation between coils is 0.19 and the minimum distance between the coils and the plasma is 0.21.
The rotational transform profile is approximately constant of $\iota \simeq 0.42$ with a mirror ratio $\Delta=0.013$.
The normalized field error on the boundary is at most 1.4\%, leading to a good agreement between the fixed boundary VMEC equilibria and the Poincaré sections, as evidenced in \cref{fig:QA_helical_ncoils2_params}.
Due to the target rotational transform of 0.41 and a total quasisymmetry residual of $0.15$ using \cref{eq:fqs}, the loss fraction of 3500 alpha particles launched from $s=0.25$ at $0.01$s is 21\%.
The currents of the helical coils are symmetric.
This configuration has a positive integrated magnetic well of 0.054, which is favorable for the stability of interchange modes.

\subsection{One Helical Coil on an Arbitrary Winding Surface}
\label{sec:helicalarbitrary}

\begin{figure}
    \centering
    \includegraphics[clip,trim=0.0cm 0.0cm 0.0cm 0.0cm,width=.42\textwidth]{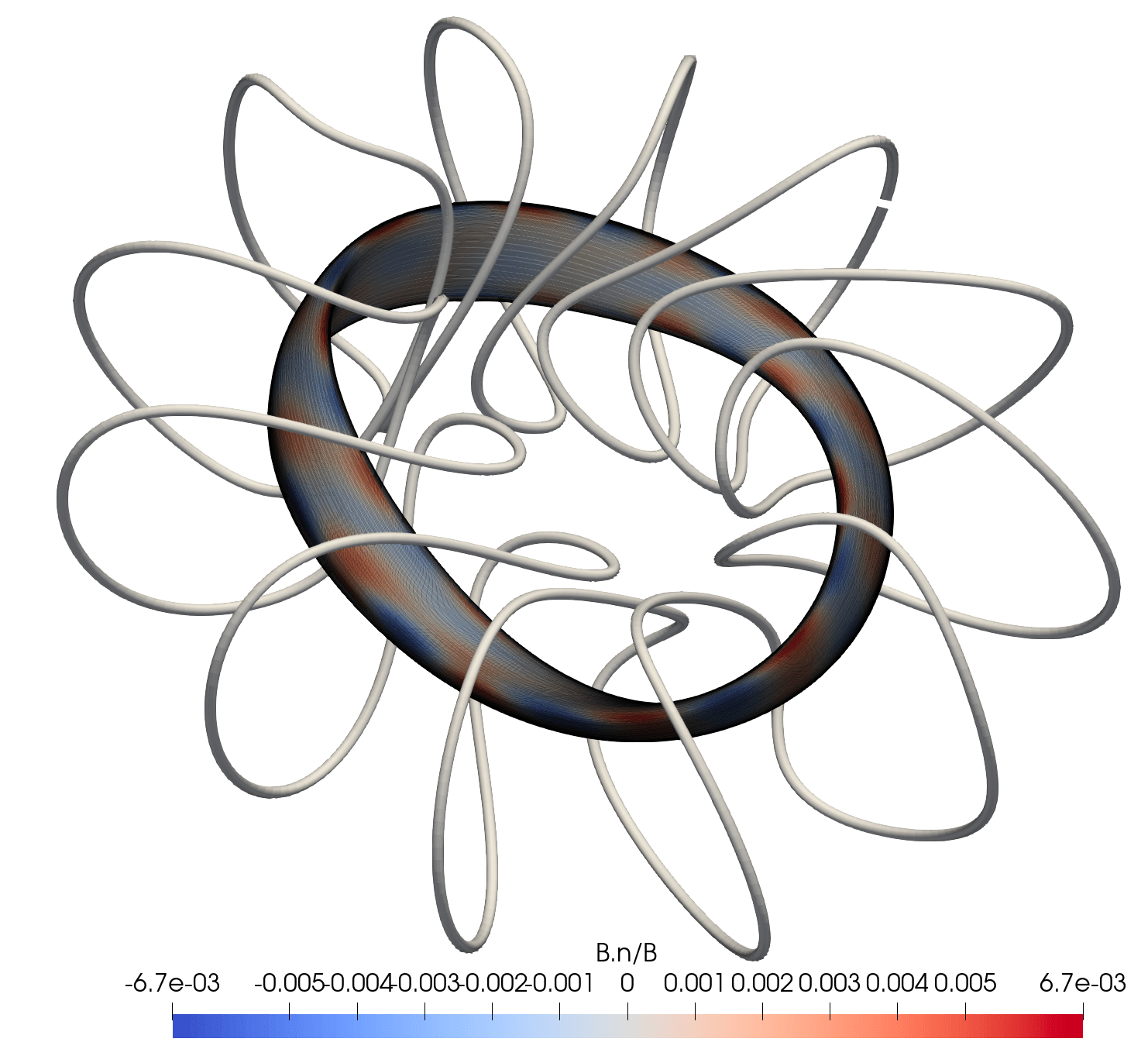}\\
    \includegraphics[clip,trim=4.5cm 0.8cm 12.2cm 6.2cm,width=.16\textwidth]{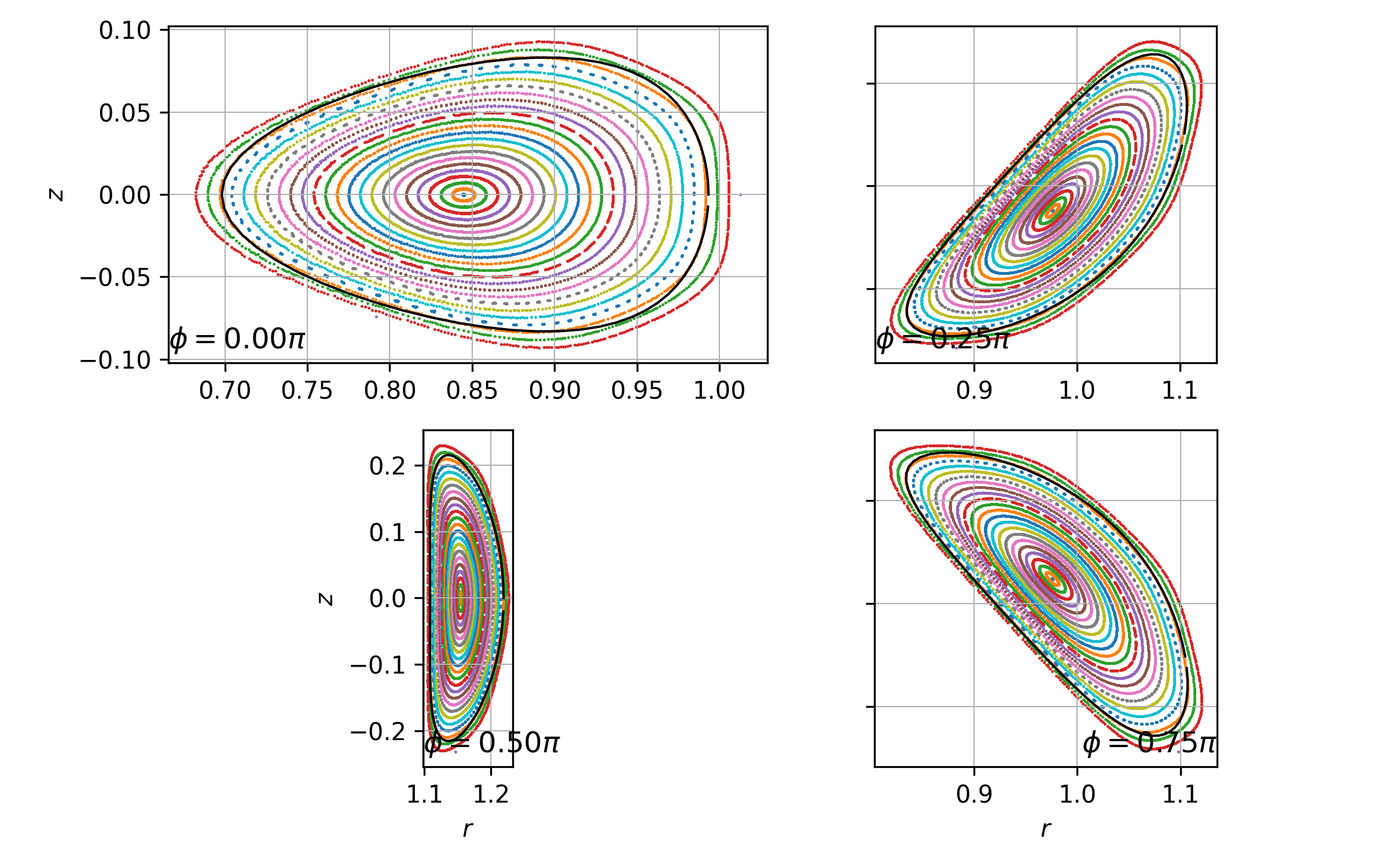}
    \includegraphics[clip,trim=0.0cm 0.0cm 0.0cm 0.0cm,width=.31\textwidth]{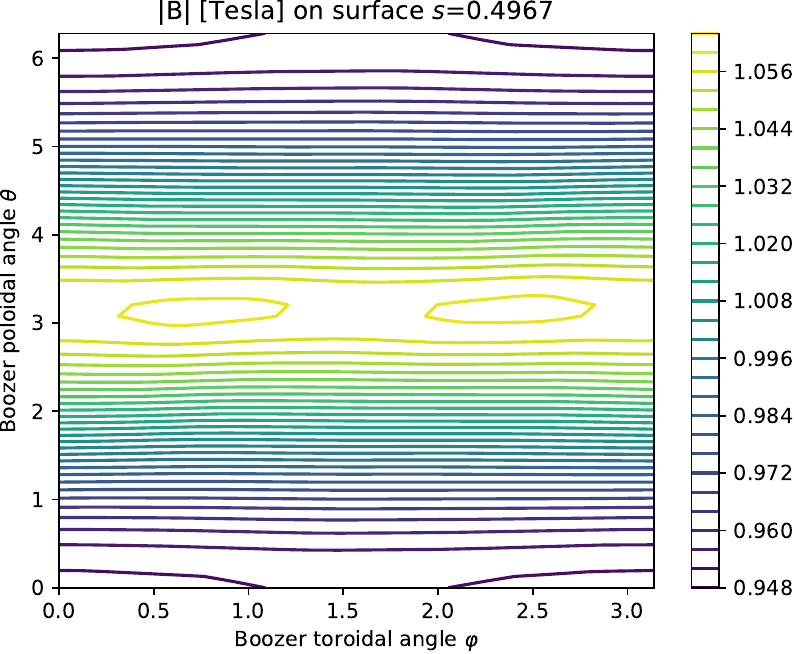}\\
    \includegraphics[clip,trim=0.1cm 11.9cm 22.8cm 0.7cm,width=.36\textwidth]{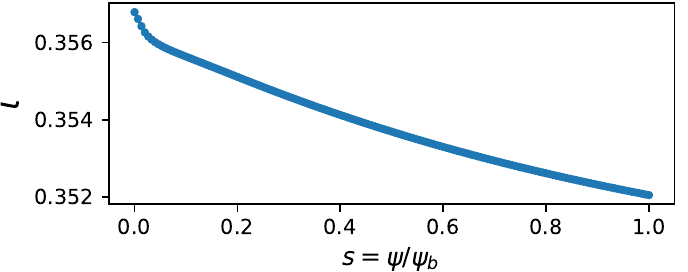}
    \caption{
    Two-field period quasi-axisymmetric stellarator with one helical coil. Top: three-dimensional view of the fixed boundary magnetic field, the normalized error coil field $\mathbf B \cdot \mathbf n/B$, and magnetic field lines in black. Upper middle: Contours of constant magnetic field in Boozer coordinates at $s=0.5$. Lower middle: Field line tracing at $\phi=0.2\pi$ and the plasma boundary (in black). Bottom: rotational transform.
    }
    \label{fig:QA_helical_ncoils1_params}
\end{figure}

We show in \cref{fig:QA_helical_ncoils1_params} a 2-field period quasi-axisymmetric stellarator with 1 helical coil where the fixed boundary equilibrium has an aspect ratio of $A=9.1$.
The coil has a total length of 57.0, a maximum curvature of 6.6, a mean-squared curvature of 3.8, and twenty Fourier modes per coil, per coordinate.
The minimum distance between the coils and the plasma is 0.33.
The rotational transform profile is approximately constant of $\iota \simeq 0.354$ with a mirror ratio $\Delta=0.005$.
The normalized field error on the boundary is at most $6.7\times10^{-3}$, leading to a good agreement between the fixed boundary VMEC equilibria and the Poincaré sections, as evidenced in \cref{fig:QA_helical_ncoils1_params}.
Due to the low to medium rotational transform and a root of the quasisymmetry error of $2.4\times10^{-5}$, the loss fraction of 3500 alpha particles launched from $s=0.25$ at $0.01$s is approximately 12\%.

\begin{figure}
    \centering
    \includegraphics[clip,trim=0.0cm 0.0cm 0.0cm 0.0cm,width=.42\textwidth]{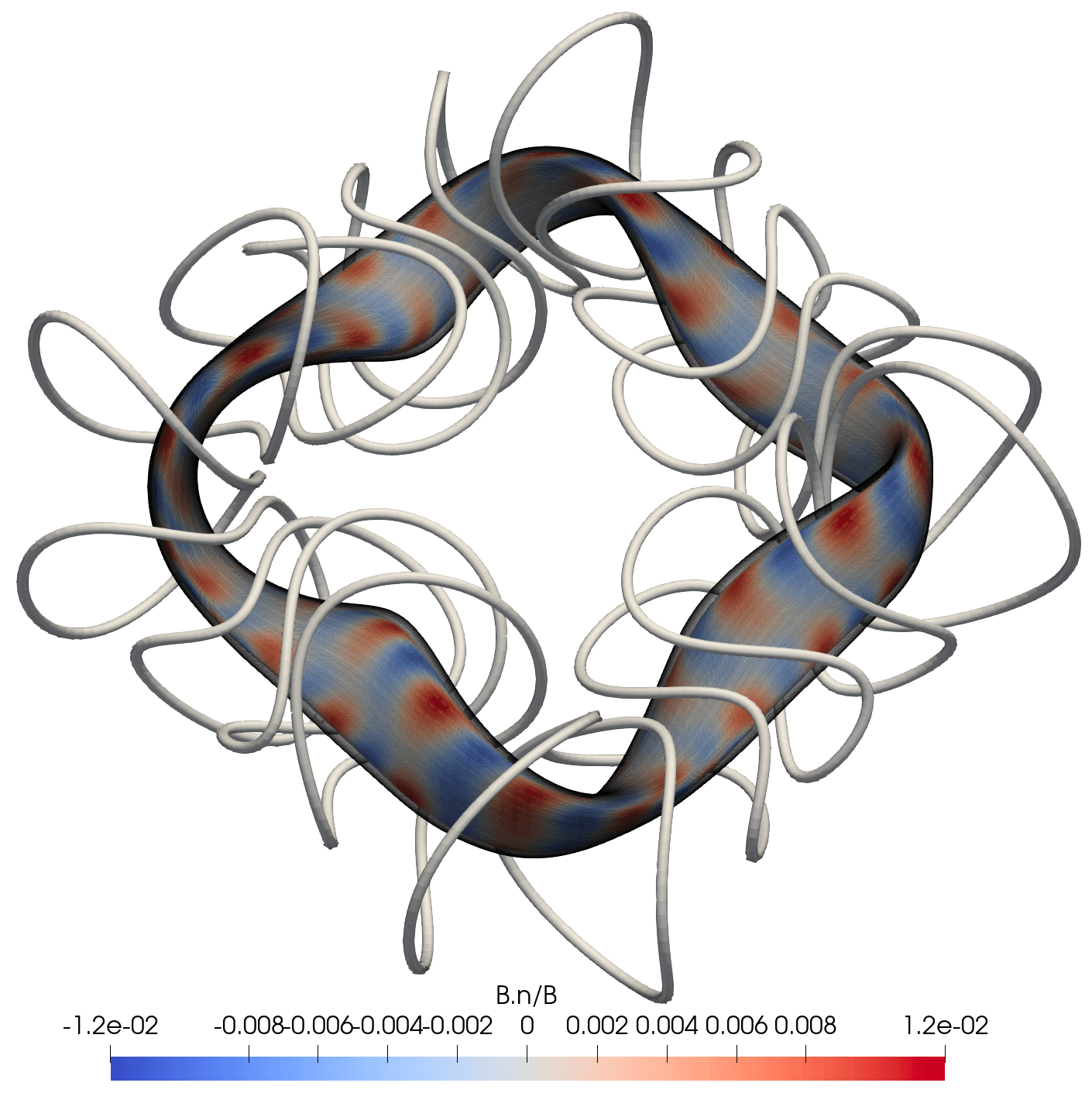}\\
    \includegraphics[clip,trim=4.3cm 6.8cm 12.6cm 0.2cm,width=.16\textwidth]{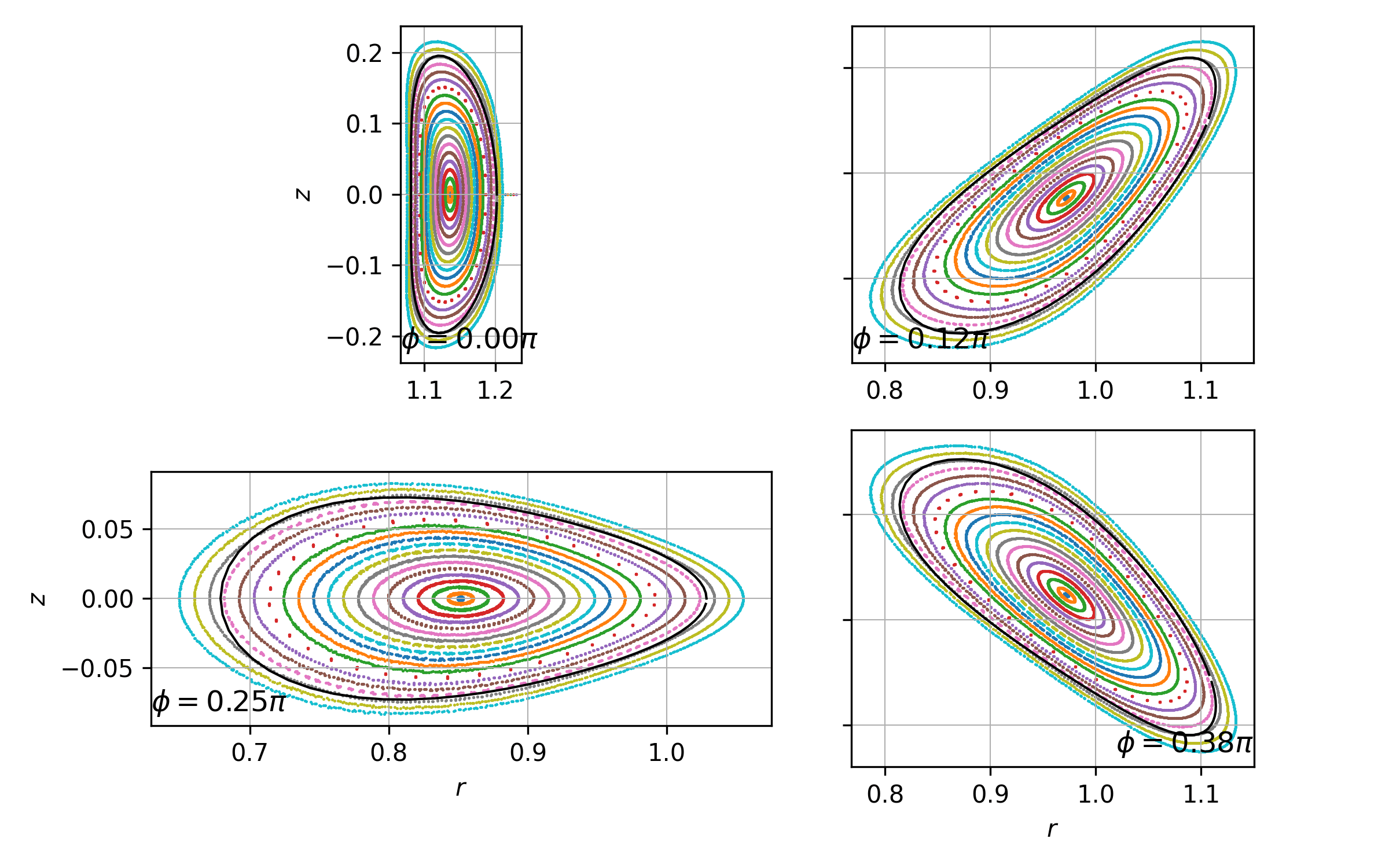}
    \includegraphics[clip,trim=0.0cm 0.0cm 0.0cm 0.0cm,width=.31\textwidth]{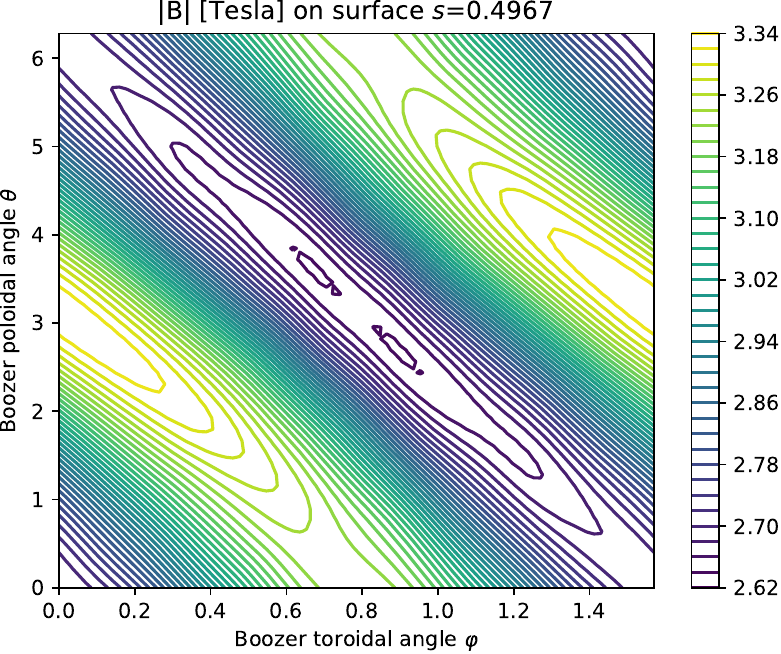}\\
    \includegraphics[clip,trim=0.1cm 11.9cm 22.8cm 0.7cm,width=.36\textwidth]{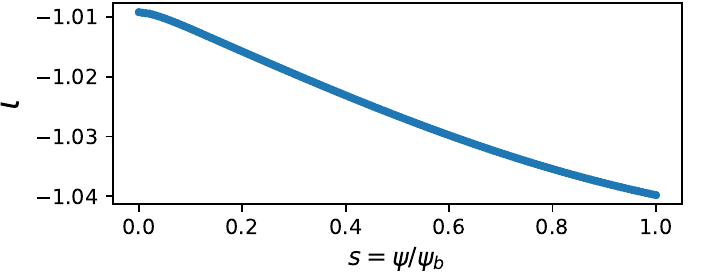}
    \caption{
    Four-field period quasi-helically symmetric stellarator with one helical coil. Top: three-dimensional view of the fixed boundary magnetic field, the normalized error coil field $\mathbf B \cdot \mathbf n/B$, and magnetic field lines in black. Upper middle: Contours of constant magnetic field in Boozer coordinates at $s=0.5$. Lower middle: Field line tracing at $\phi=0.\pi$ and the plasma boundary (in black). Bottom: rotational transform.
    }
    \label{fig:QH_helical_ncoils1_params}
\end{figure}

We show in \cref{fig:QH_helical_ncoils1_params} a 4-field period quasi-helically symmetric stellarator with 1 helical coil where the fixed boundary equilibrium has an aspect ratio of $A=8.7$.
The coil has a total length of 60.0, a maximum curvature of 9.2, a mean-squared curvature of 10.1, and eighteen Fourier modes per coil, per coordinate.
The minimum distance between the coils and the plasma is 0.20.
%
% \ag{i think there is a typo in the rotational transform here, shouldn't it be 1.025?}
% Great, thanks!
The rotational transform profile is approximately constant of $\iota \simeq 1.025$ with a mirror ratio $\Delta=0.01$.
The normalized field error on the boundary is at most $1.2\times10^{-2}$, leading to a good agreement between the fixed boundary VMEC equilibria and the Poincaré sections, as evidenced in \cref{fig:QH_helical_ncoils1_params}.
Due to the high rotational transform and a root of the quasisymmetry error of $2.1\times10^{-2}$, the loss fraction of 3500 alpha particles launched from $s=0.25$ at $0.01$s is approximately 0.14\%.

\section{\label{sec:QAQHflexible} Flexible QA-QH Stellarators with a Single Set of Coils}

\begin{figure}
    \centering
    \subfloat[]{\includegraphics[clip,trim=0.1cm 2.1cm 2.0cm 2.3cm,width=0.45\textwidth]{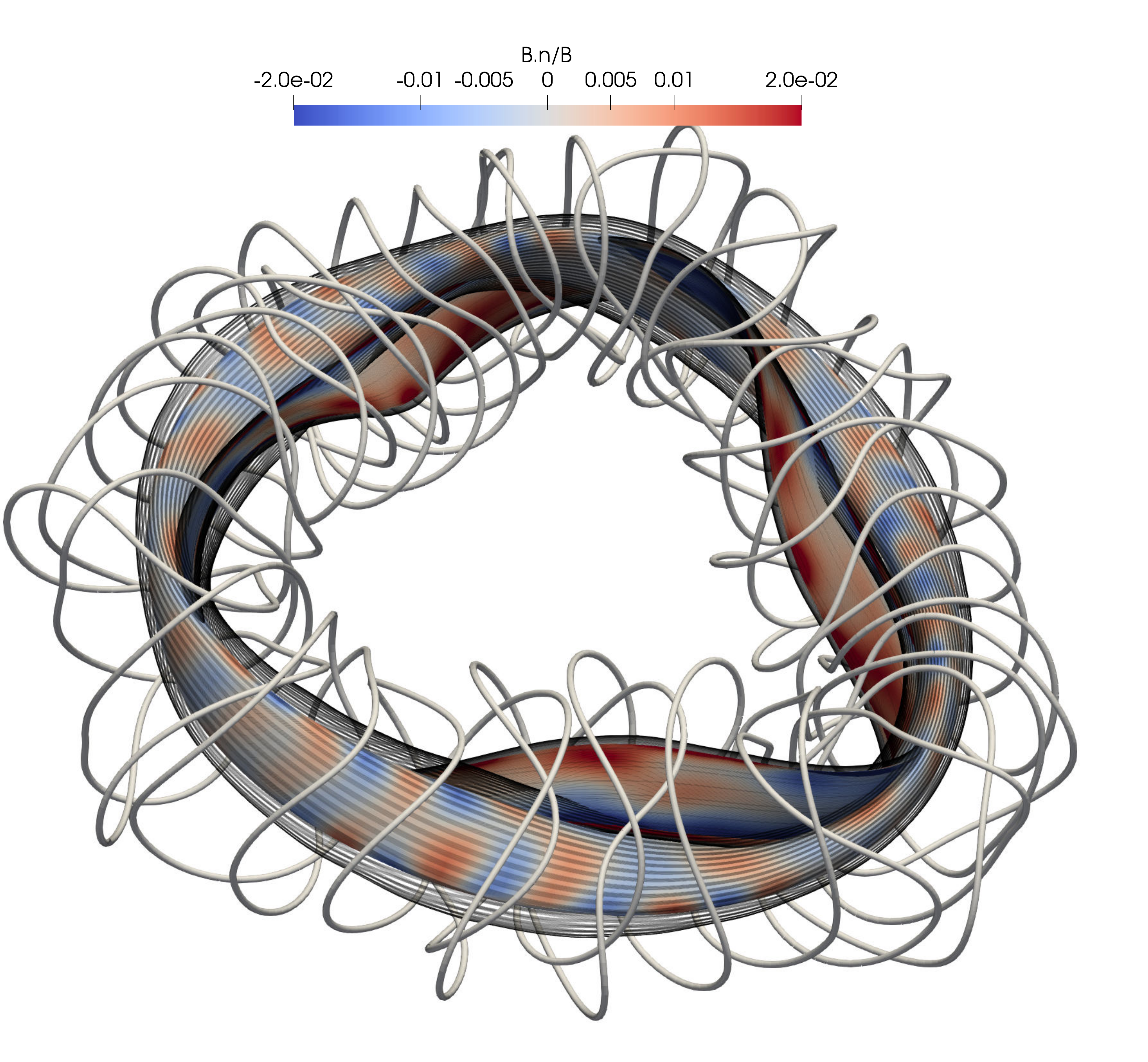}}
    \hfill
    \vspace{-0.4cm}
    \subfloat[]{\includegraphics[clip,trim=0.0cm 0.0cm 0.0cm 0.0cm,width=0.235\textwidth]{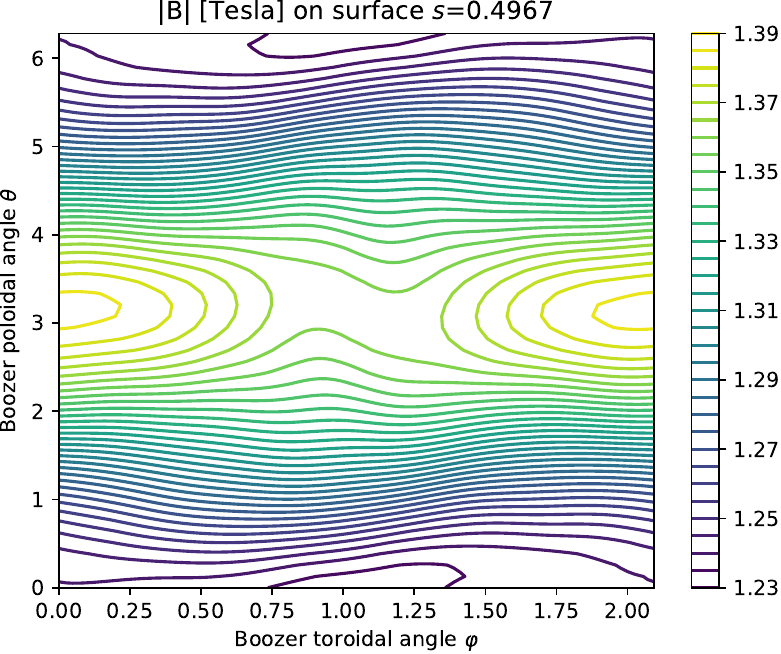}}
    \hfill
    \vspace{-0.4cm}
    \subfloat[]{\includegraphics[clip,trim=0.0cm 0.0cm 0.0cm 0.0cm,width=0.235\textwidth]{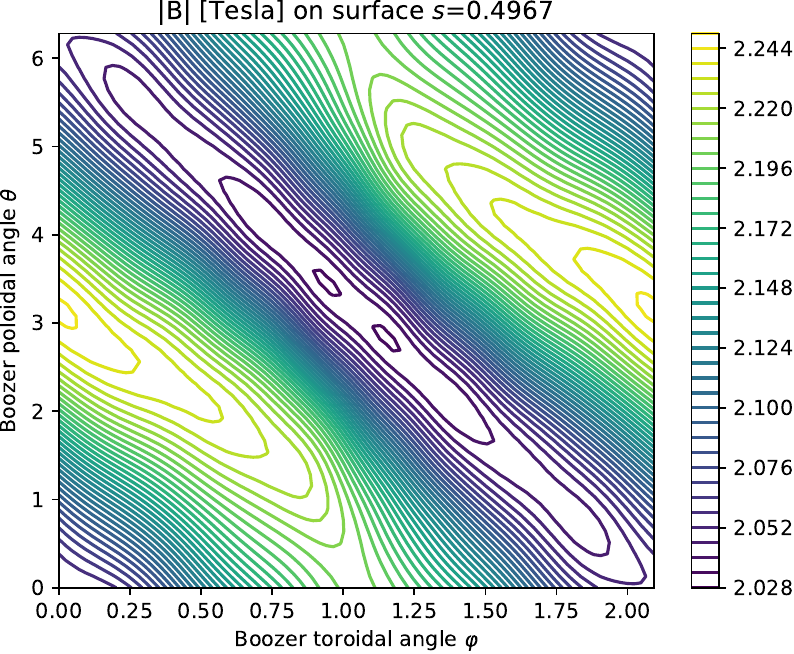}}

    \subfloat[]{\includegraphics[clip,trim=4.7cm 6.4cm 12.3cm 0.3cm,width=0.117\textwidth]{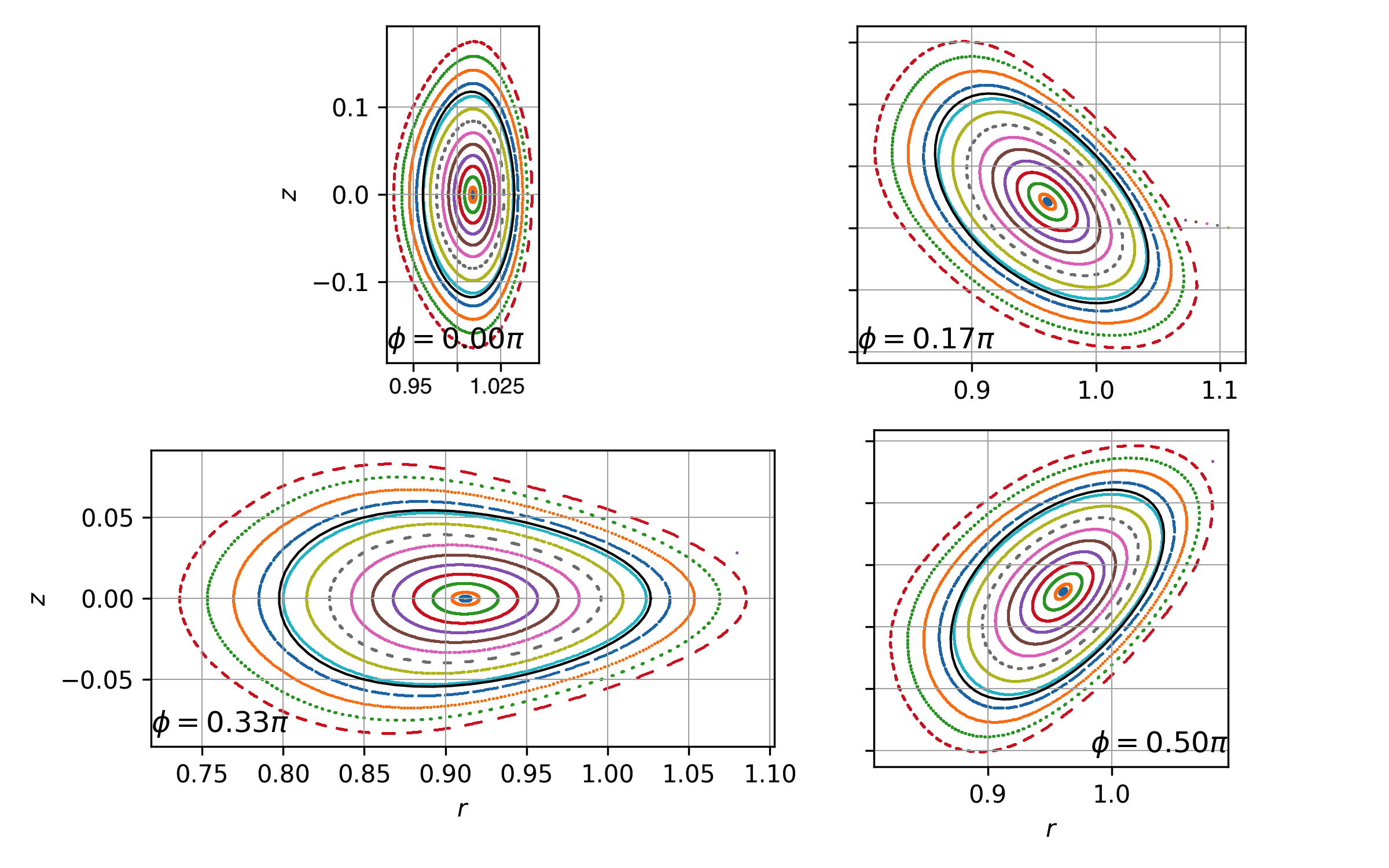}}
    % \hfill
    \subfloat[]{\includegraphics[clip,trim=5.5cm 6.6cm 12.6cm 0.3cm,width=0.083\textwidth]{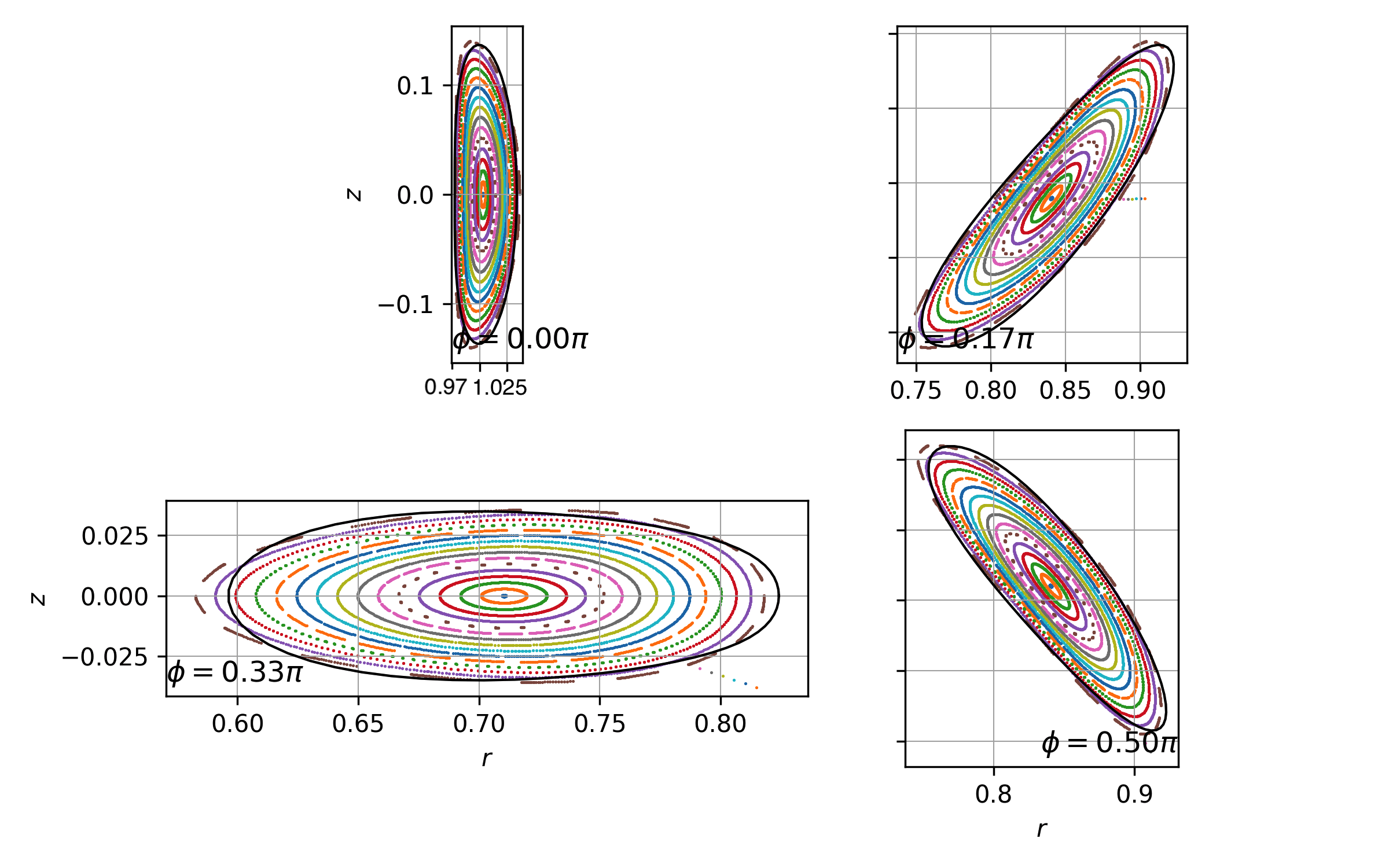}}
    \hfill
    \subfloat[]{\begin{minipage}{0.275\textwidth}
        \centering
        \vspace{-3.6cm}
        \includegraphics[clip,trim=0.1cm 12.0cm 22.8cm 0.7cm,width=\textwidth]{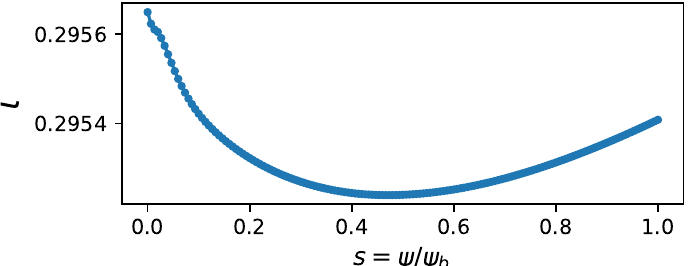}\\
        \includegraphics[clip,trim=0.1cm 12.0cm 22.9cm 0.7cm,width=\textwidth]{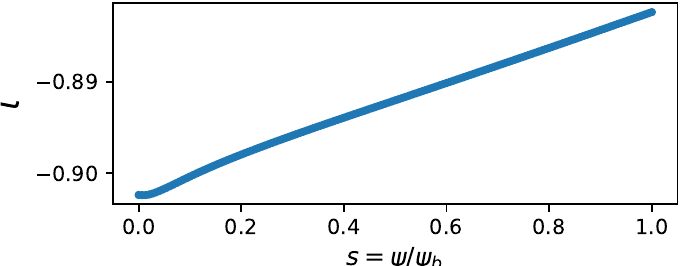}
    \end{minipage}}\vspace{-0.2cm}
    \caption{Three-field period flexible stellarator with seven non-planar coils per half field-period. (a) Three-dimensional view of the fixed boundary magnetic field, the normalized error coil field $\mathbf B \cdot \mathbf n/B$, and magnetic field lines in black. (b, c) Contours of a constant magnetic field in Boozer coordinates at $s=0.5$, QA and QH. (d, e) Field line tracing at $\phi=0.0\pi$ and the plasma boundary, for QA and QH. (f) Rotational transform, $\iota$, for QA (top) and QH (bottom).\vspace{-0.6cm}}
    \label{fig:QA_QH_flexible_params}
\end{figure}

We show in \cref{fig:QA_QH_flexible_params} the resulting coils and magnetic field equilibria for the 3-field period flexible quasi-axisymmetric (QA) with $A=12.2$ and quasi-helically (QH) symmetric with $A=13.2$ configuration with 7 modular coils per half field-period.
The coils have a total length of 2.2 each, a maximum curvature of 7.0, 7.3, 6.3, 7.4, 6.1, 5.7, and 5.4, a mean-squared curvature of 14.1, 14.0, 14.1, 14.0, 14.3, 14.2, and 14.1, and five Fourier modes per coil, per coordinate.
The minimum separation between coils is 0.08 and the minimum distance between the coils and the plasma is 0.10 for QA and 0.03 for QH.
The rotational transform profile is approximately constant of $\iota \simeq 0.295$ for the QA configuration with a mirror ratio $\Delta=0.006$, and $\iota \simeq 0.98$ for the QH configuration with a mirror ratio $\Delta=0.007$.
The normalized field error on the boundary is at most 2.0\%, leading to a good agreement between the fixed boundary VMEC equilibria and the Poincaré sections, as evidenced in \cref{fig:QA_QH_flexible_params}.
Due to the target rotational transform of 0.3 and a total quasisymmetry residual of $0.16$ using \cref{eq:fqs}, the loss fraction of 3500 alpha particles launched from $s=0.25$ at $0.01$s for the QA configuration is 18\%.
The QH configuration has a quasisymmetry residual of $0.007$ and a loss fraction of 2.6\%.
The currents of the coils for each configuration are listed in \cref{tab:QAQHcurrents}.
This QA configuration has a positive integrated magnetic well of 0.012, while the QH configuration has a negative integrated magnetic hill of -0.02.

% \begin{figure}
%     \centering
%     \includegraphics[clip,trim=0.4cm 0.6cm 1.1cm 1.8cm,width=.45\textwidth]{dual_coils_QA_currents.png}
%     \includegraphics[clip,trim=0.4cm 0.6cm 1.1cm 1.8cm,width=.45\textwidth]{dual_coils_QH_currents.png}
%     \caption{Caption}
%     \label{fig:QA_helical_ncoils2_3D_currents}
% \end{figure}

\begin{table}[htbp]
    \centering
    \caption{Coil currents for the flexible QA-QH stellarator with 7 coils per half field-period normalized by the maximum current of the QA configuration.}
    \begin{tabular}{c|c|c}
        \toprule
        Coil $i$ & QA currents $I^{QA}_i$ & QH currents $I^{QH}_i$ \\
        \midrule
        1 & 1.0000 & 0.3085 \\
        2 & -0.1399 & 0.5549 \\
        3 & 0.5439 & -0.3409 \\
        4 & 0.4491 & 0.8404 \\
        5 & 0.9590 & 0.8295 \\
        6 & -0.2126 & -0.0239 \\
        7 & 0.7764 & 0.5834 \\
        \bottomrule
    \end{tabular}
    \label{tab:QAQHcurrents}
\end{table}

\section{\label{sec:conclusions} Conclusions}

In this work, single-stage optimization was used to find sets of coils for the following configurations: simplified stellarators with six to eight coils, quasisymmetric stellarators with one to three independent coils, external coils that are not linked with the plasma, helical coils, and flexible configurations with one set of coils that produces both a quasi-axisymmetric and a quasi-helical symmetric configuration.
It is found that, although quasi-axisymmetric configurations allow for simpler stellarators with a lower number of coils, the level of quasisymmetry and aspect ratio necessary to achieve small loss fractions is more stringent than quasi-helically symmetric configurations.
Furthermore, we see that flexible stellarators with multiple configurations and a single set of seven non-planar coils per half field-period are possible.
However, the coil currents need to be reversed when changing configurations which may pose engineering challenges.
A study of the trade-offs between stellarator flexibility and coil complexity will be performed in future studies.

We note that, while different applications of single-stage optimization were pursued here, many optimizations can still be performed using such a method, for which we enumerate a few.
The single-stage optimizations carried out here using a fixed boundary equilibrium show the presence of a set of nested flux surfaces outside the equilibrium.
Therefore, as an additional step, a QFM surface can be created from the obtained set of coils as the initial condition for a more refined optimization study.
The flexibility studies shown here have solely used coil currents as possible degrees of freedom.
However, using the same space curves, small coil rotations and displacement could be used to improve flexibility.
Additional planar coils, windowpane coils, or trim coils, could be added to further increase flexibility similar to what was proposed in Refs. \!\![\onlinecite{Risse2011},\onlinecite{Wiedman2023}].
The addition of permanent magnets \cite{Zhu2020,Madeira2024} can also be used in a single-stage framework to simplify coils or improve flexibility.
Finally, we mention that finite plasma $\beta$ effects were not included, and additional metrics such as MHD stability and magnetic shear can be added in future work.

\vspace{-0.5cm}
\section{Acknowledgments}
\vspace{-0.1cm}

We acknowledge the effort by the SIMSOPT team to develop the tools used in this work.
JL would like to acknowledge Chris Smiet, Erol Balkovic, Jason Smoniewski, and Allan Goodman for help with the SIMSOPT code.
This work has been carried out within the framework of the EUROfusion Consortium, via the Euratom Research and Training Programme (Grant Agreement No 101052200 — EUROfusion) and funded by the Swiss State Secretariat for Education, Research and Innovation (SERI). Views and opinions expressed are however those of the author(s) only and do not necessarily reflect those of the European Union, the European Commission, or SERI. Neither the European Union nor the European Commission nor SERI can be held responsible for them. This research was also supported by a grant from the Simons Foundation (1013657, JL).
This research used resources of the National Energy Research
Scientific Computing Center, a DOE Office of Science User Facility supported by the Office of Science of the U.S. Department of Energy under Contract No. DE-AC02-05CH11231 using NERSC award NERSC DDR-ERCAP0030134.
This work used Jetstream2 at Indiana University through allocation PHY240054 from the Advanced Cyberinfrastructure Coordination Ecosystem: Services \& Support (ACCESS) program, which is supported by National Science Foundation grants \#2138259, \#2138286, \#2138307, \#2137603, and \#2138296.

\bibliography{references}

\end{document}